\documentclass{article}
\usepackage{graphicx}
\usepackage{authblk}
\begin{document}
\title{Maximum mass and radius of strange stars in Finch-Skea geometry in dimensions $D\geq4$.}
\author[1]{B. Das}
\author[2] {K. B. Goswami}
\author[3] {A. Saha}
\author[4]{P. K. Chattopadhyay\thanks{Corresponding Author, E-mail: pkc$_{-}$76@rediffmail.com}}

\affil[1,2,3,4]{Department of Physics, Coochbehar Panchanan Barma University\\
 Vivekananda Street, District: Coochbehar\\ Pin: 736101, West Bengal, India}
\affil[3]{Department of Physics, Alipurduar College, Alipurduar, Pin:736122, West Bengal, India}

\date{Received: date / Accepted: date}
\maketitle

\begin{abstract}
In this article, we demonstrated a stellar model for compact star in presence of strange matter embedded in $D\ge4$ dimensional space-time defind by Finch-Skea metric. To study the relevant physical properties of the interior matter, we consider the equation of state $(henceforth~EOS)$ as proposed in MIT bag model given by $p=\frac{1}{3}(\rho-4B)$, where $B$ is termed as bag constant. The Mass-Radius relationships in four and higher dimensions are determined using the range of values of surface density through the relation $\rho_{s}=4B$ for which bulk strange matter may be a viable issue for compact objects. Here we choose the range of $B$ such that stable strange matter may exist at zero external pressure relative to neutron. We note that a maximum value of the stellar radius is exist when $B$ is fixed at a given allowed value for which metric functions considered here to be real. This is the maximum allowed radius $(b_{max})$ in this model which depends on surface density of a strange star. In four dimensions the compactness of a star is found to be greater than 0.33. In case of higher dimensions ($D>4$), we observed different values of compactness. Causality conditions are satisfied interior to the star upto maximum allowed radius $(b_{max})$ for which metric function is real. The validity of energy conditions, surface red-shift and other parameters of the stellar configuration are studied and found new results. Stability of the system is also studied. 
\end{abstract}


\section{Introduction}
Compact objects in relativistic astrophysics are mysterious type of objects whose matter density is extremely high and are formed due to gravitational collapse of matter when internal nuclear fuel get exhausted to maintain hydrostatic equilibrium. Depending on the initial mass of a main sequence star, its end state forms different compact objects having wide range of compactness (ratio of mass to radius) factor. More is the compactness more is the dense packing of matter interior of a compact object. Typically known compact objects are White Dwarfs, Neutron Stars and Black Holes. The core of more massive stars has the possibility to converted into neutron stars or black holes in the form of supernovae explosion. As the density of such stars is enormously high ($\sim$ nuclear matter density or above), it is interesting to analyze the physical properties of matter at these ultra-high densities regime. Accordingly considerable interest has been grown among the researchers to study the properties of such astrophysical objects in the past couple of years in high energy physics and relativistic astrophysics. Present day observational data from different space based satellites show that a large number of such astrophysical objects may exist in our universe having widely varying masses and radii. Lots of investigations has been carried out by several researchers to predict observed mass-radius relation of such objects from theoretical standpoints. However, as the interior composition at these high densities is quite unknown, there is no unique theory using which it may be possible to predict the mass-radius relation of all compact objects. In case of neutron star, the gravitational collapse is counter balanced by the repulsive degeneracy pressure of neutrons. If the gravitational attraction is strong enough due to higher initial mass to overcome the repulsive degeneracy pressure of neutrons, the star again start to shrink further into a lower radius and higher density than a conventional neutron star can hold. The matter densities of such stars are normally higher than the density of nuclear matter. However, maximum mass and radius of such stars have lower values than that of neutron stars and their mass to radius ratio i.e. compactness factor is higher than that of neutron stars. This may be explain as follows: due to inward strong gravitational pull hadrons phase may converted into quark phase and as quarks are smaller in size than hadrons, packing becomes more. Hence more mass can be accommodated within smaller radius. The most general assumption for such high densities is that nucleons may converted into hyperons or either form condensates. However, the prediction of estimated masses and radii of such compact objects could not be possible in a good precession using the standard theoretical models available for the analysis of the neutron stars. Such as the the prediction of physical properties of many compact stars like Her X-1, Vela X-1, 4U 1538-52, SAX J 1808.4-3658 and 4U 1728-34 and PSR J1614-2230 using the standard models for neutron star are not in well agreement with observed data. Such disagreement between theoretical prediction and observed data forced many researchers to think about the composition of interior matter of such compact stars by introducing new concept. Witten \cite{Witten} first proposed that strange quark matter may exist interior of a compact object which is based on the study using MIT bag model. The study indicates that quarks may be free inside a compact objects and form relativistic Fermi gas and might not be in a confined hadronic phase. These quarks may form a larger colourless region having equal proportion of Up $(u)$, Down $(d)$ and Strange quarks $(s)$ at such ultra-high densities. Accordingly, exact composition of matter especially near the core region of compact stars is an open area to be explore in relativistic astrophysics till date. In the GR framework, several authors have successfully used the MIT bag EOS to describe
strange quark stars \cite{MB,LP,JDV,GL,SRC}. To describe the observational aspects of many compact objects $(neutron/strange~stars)$, it has been predicted that strange quarks might exist \cite{Glendenning,Bombaci} in the interior along with other particles. These objects may be grouped into new category coined as the strange star ($SS$) candidates. It requires a high temperature environment or baryon chemical potential to convert hadronic matter into quark-gluon plasma, which may be achieved for a short time scale in ultra-relativistic heavy-ion collisions \cite{Madsen}. Quark phase inside a compact object may achieve in several processes out of which two possible distinct ways \cite{Witten,Itoh,Bodmer} are: (i) hadrons may be converted into quark phase at the time of the creation of early universe when temperature was extremely high, and (ii) due to tremendous inward gravitational pull, neutron stars may be converted into strange stars at very high densities. So far the stability is concerned, quark matter composed of two flavoured quarks such as Up ($u$) and down ($d$) is likely to be unstable. After the prediction of Witten \cite{Witten} and Bodmer \cite{Bodmer} that quark matter in true ground state may be of strange quark matter, the concept of inclusion of Strange quarks ($s$) into the two flavoured system effectively reduces the energy per baryon and hence quark matter composed of the three-flavoured quarks ($u$, $d$ and $s$) becomes more stable as compared to two flavoured-system ($u$ and $d$ quarks). It is considered in the theory of MIT bag model that the quark matter is made up with $u$ and $d$ quarks, $s$ quarks and electrons. Masses of $u$ and $d$ quarks are zero whereas $s$ quarks have non-zero mass. This collection of quarks and electrons may exist in a region characterised by vacuum energy density $B$ (also known bag constant)  and are considered to be degenerate Fermi gases. If it is assumed that quarks are non-interacting and massless, then the relation between the quark pressure $p_q$ and quark density $\rho_q$ is given as $p_{q}=\rho_{q}/3$. In presence of $B$, the total energy density and total pressure of the quark assembly are $\rho=\rho_{q}+B$ and $p=p_{q}-B$ respectively. Kapusta \cite{Kapusta} derived the EOS for quark matter composed of strange quarks is given below:
\begin{equation}
p_{r}=\frac{1}{3}(\rho-4B),\label{eq00}
\end{equation}
Madsen \cite{Madsen} evaluated the minimum value of $B$ considering the value of strong fine structure constant $\alpha_{s}=0$ at zero external pressure. This minimum value of $B$ should be less than the mass of neutron ($939.6~MeV$) or $B^{1/4} < (145.9~MeV)$ \cite{Madsen}. On the other hand this limit will be $B^{1/4} < (144.4~MeV)$ \cite{Madsen}, if one consider the stability of quark matter relative to iron. In our present work we considered the lower limit of $B$ as $B^{1/4}_{min}= 145.9~MeV (\equiv B=57.55~MeV/fm^3)$ \cite{Madsen,Kapusta}. This value of $B$ is the lowest allowed value necessary for the stability of atomic nuclei against decay into non-strange quark which is in general not observed. To get an upper limit of $B$ relative to iron for stability of strange quark matter, Madsen \cite{Madsen} pointed out that this value is about $B^{1/4}_{max}= 162.8 MeV$. With respect to neutron this value becomes  $B^{1/4} < 164.4~MeV$ \cite{Madsen}. The above limits are evaluated considering zero external pressure as well as neglecting the mutual interaction among the constituent quarks \cite{Madsen,Kapusta}. For our present model we shall consider the upper limit of $B$ as $B^{1/4} < 164.4~MeV (\equiv 95.11~MeV/fm^3)$. Therefore from the study of Madsen \cite{Madsen}, one may consider that so far the stability of bulk quark matter is concerned, the value of $B$ should lie in the range $57.55$ $MeV/fm^3$ $< B <$ $95.11~MeV/fm^3$ relative to neutron. However, all the above limits of $B$ get reduced for non-zero fine structure constant $\alpha_{s}$ \cite{Farhi}. 

It is expected that unequal principal stresses may arise inside the interior of compact objects when its matter densities are above the nuclear matter density. Unequal principal stresses means the radial $(p_{r})$ and tangential $(p_{t})$ pressure are unequal at any internal point except at the centre of the star. In the Newtonian regime the prediction of existence of pressure anisotropy was proposed first by Jeans \cite{JH} for self-gravitating objects. In the context of general relativity Lemaitre \cite{LG} considered that anisotropy in pressure as the local effect and using this concept Lamaitre predicted that the surface gravitational potential may have higher value in presence of pressure anisotropy than isotropic one $(p_r=p_t)$. In case of compact star the nuclear interactions are supposed to be relativistic in nature when matter density $(\rho)$ exceeds $10^{15}gm/cm^3$ and such stars may likely to be anisotropic in nature \cite{MR}. In the context of Newtonian and general relativity, Herrera \cite{Herrera1} predicted that in case of self gravitating system local anisotropy may appear in the interior of compact object in the high density regime. Subsequently, many investigators have been interested to find out the new exact solutions of Einstein field equations incorporating anisotropy in the theory. Apart from that anisotropy may arise from other physical phenomenon. Such as the anisotropic distribution of pressure in high density environment \cite{Kippenhahn} may be explain in presence of super fluid of type 3A or a solid core. Anisotropy may arise from the condensation pion \cite{Sawyer} and and also from the different type of phase transitions \cite{Sokolov}.
 
To analyse some issues both in astrophysics and cosmology, conventional approach of four dimensions may not be sufficient to explain all the physical features of some astrophysical or cosmological phenomenon. It is sometime required to include extra dimensions in the theory to explain these phenomenon. Such as in high energy physics, it is often required greater than conventional four dimensions for their consistent formulation of important theories of interactions among constituent particles. Accordingly, it is important to study the generalization of four dimensional approach into higher dimensional theories. In this context to unify gravity with electro-magnetic interaction, \cite{Kaluza} and \cite{Klein} were the first who introduced an extra dimension in their theory independently and studied the effects of incorporation of extra dimension. Similarly the generalization of the results of GTR in conventional four dimensions is essential in the context of higher dimensional theory. To describe the structure of present universe satisfactorily, Detweilar \cite{Detweiler} built some cosmological model in higher dimensional approach which may undergo a spontaneous compactification of space-time. This corresponds to a product space represented by the term $M^4 \times M^d$ where $M^d$ is the compact inner space. In the five dimensional anti-de Sitter space-time ($AdS_{5}$), Randall and Sundram \cite{Randall} predicted that the four dimensional aspect of Newtonian gravity might be recovered in the low energy regime which is an interesting signature of general relativity. Although Randall and Sundram \cite{Randall} considered that the extra dimension is not compact in their theory. Many investigators have successfully generalized the four dimensional results of astrophysics and cosmology in the higher dimensional theories. Such as spherically symmetric Schwarzschild and Reissner-N{\"o}rdstrom blackholes \cite{Chodos,Gibbons}, Kerr black holes \cite{Myers1,Mazur,Xu}, black holes in compactified space-time \cite{Myers2}, no hair theorem \cite{Sokolowsky}, Hawking radiations \cite{Myers1}, Vaidya solution \cite{Iyer} has been studied in higher dimensions. Shen and Tan \cite{Shen} obtained a global regular solution of Schwarzschild space-time in higher dimensional context. New studies have been made to evaluate the mass to radius ratio of a uniform density star in the framework of higher dimensions \cite{Paul2}. The structure of neutron star have been studied in the literature \cite{Liddle} in the context of Kaluza-Klein theory as a consequences of extra dimensions. Using numerical approach Tikekar and Jotania \cite{Jotania} established the physical acceptibility of models for strange star admitted in the set-up with compactness higher than 0.3. In this paper, we have studied the maximum allowed radius, maximum mass and other relevant properties of strange star using MIT bag model EOS $p=\frac{1}{3}(\rho-4B)$ for the strange matter in four and higher dimensions considering Finch-Skea \cite{FS} metric. Here we use the value of $B$ within the range $57.55~MeV/fm^3$ $<$ $B$ $<$ $95.11~MeV/fm^3$ required for stable strange matter at zero external pressure relative to neutron \cite{Madsen}. The basic aim is to predict suitable model for strange stars to explain their observed properties and internal composition of matter content. 

The article is written in the following manner: In \S 2, we established the Einstein field equations in Finch-Skea geometry assuming anisotropic pressure in higher dimensional space-time. Physically viable stellar model in higher dimensions is ensured using boundary conditions at the surface of the star. In \S 3, we have studied the model numerically as the field equations are complicated in nature using the allowed values of bag constant $B$. Maximum allowed radius of strange star and corresponding maximum mass, surface red-shift are obtained in \S 4. Physical viability of this model have been studied in \S 5 through physical applications. In \S 6, we have discuss the stability conditions of the model. In \S 7 stability against the variation of Lagrangian perturbation of radial pressure with frequency is studied. We conclude by summarizing our new findings in \S 8.

\section{Field Equations \& Model Parameters}
The interior space-time of a spherically symmetric, static cold compact star is described by the following line element 
\begin{equation}
ds^{2} = -e^{2\nu(r)}dt^{2}+e^{2\mu (r)}dr^{2}+r^{2}d\Omega_{n}^{2},\label{eq01}
\end{equation}
where $\nu(r)$, $\mu(r)$ are the two metric functions and $n$ is related to space-time dimensions $D$ through the relation $n=D-2$ and $d\Omega_{n}^{2}=d\theta_{1}^{2}+sin^{2}\theta_{1}(d\theta_{2}^{2}+sin^{2}\theta_{2}d\theta_{3}^{2}+sin^{2}\theta_{2}sin^{2}\theta_{3}d\theta_{4}^{2}+.....+sin^{2}\theta_{2}sin^{2}\theta_{3}...sin^{2}\theta_{n-1}d\theta_{n}^{2})$ is the angular part of the metric on n-sphere.\\ 
The energy-momentum tensor in D-dimensional space-time with fluid pressure anisotropic in nature is given by the general form,
\begin{equation}
T_{ij} = \mbox{diag}~(-\rho, p_{r}, p_{t}, p_{t},......p_{t}),\label{eq02}
\end{equation}
where $\rho$, $p_{r}$ and $ p_{t}$ are known as the energy density, radial and tangential pressure respectively.
The Einstein field equation in D-dimensional space-time is given by,
\begin{equation}
{\bf R}_{ij}-\frac{1}{2}g_{ij}{\bf R} = \frac{8\pi G_{D}}{c^2} \; {\bf T}_{ij},\label{eq03}
\end{equation}
where $G_{D}=G~V_{D-4}$ is known as gravitational constant in $D$ dimensions, $G$ is the Newtonian constant in four dimensions and $V_{D-4}$ represent the volume of extra space given by $V_{x}=\frac{\pi^{x/2} r^{x}}{\Gamma(1+x/2)}$ where $x=D-4$. $R_{ij}$ is Ricci tensor and $R$ is the Ricci scalar. Using eqs.~(\ref{eq01}) and (\ref{eq02}) in the eq.~(\ref{eq03}), the Einstein field equations reduces to the following set of equations:
\begin{equation}
\frac{n(n-1)(1-e^{-2\mu})}{2r^{2}}+\frac{n\mu^{\prime}e^{-2\mu}}{r}= \frac{8\pi G_{D}}{c^2} \rho,\label{eq04}
\end{equation}
\begin{equation}
\frac{n\nu^{\prime}e^{-2\mu}}{r}-\frac{n(n-1)(1-e^{-2\mu})}{2r^2}= \frac{8\pi G_{D}}{c^2}p_{r},\label{eq05}
\end{equation}
\[
  e^{-2\mu}\left[\nu ^{\prime \prime }+{\nu ^{\prime }}^{2}-\nu ^{\prime
}\mu ^{\prime }-\frac{(n-1)(\mu ^{\prime}-\nu ^{\prime })}{r}\right] 
\]
\begin{equation}
 -\frac{(n-1)(n-2)(1-e^{-2\mu})}{2r^2}
=\frac{8\pi G_{D}}{c^2}p_{t}. \label{eq06}
\end{equation}
Using eqs. (\ref{eq05}) and (\ref{eq06}), we get the following equation,
\[
 e^{-2\mu}\left[\nu ^{\prime \prime }+{\nu ^{\prime }}^{2}-\nu ^{\prime
}\mu ^{\prime }-\frac{\nu ^{\prime }}{r} \right.
\]

\begin{equation} 
\left. - \frac{(n-1)\mu ^{\prime}}{r}-\frac{(n-1)(n-2)(1-e^{2\mu})}{2r^2}\right]
= \frac{8\pi G_{D}}{c^2}\Delta. \label{eq07}
\end{equation}
Here we define the difference between radial and transverse pressure at any radial distance $r$ as the measure of pressure anisotropy inside the star and denoted by $\Delta=p_{t}-p_{r}$. Clearly the anisotropy depends on the metric potentials $\nu(r)$ and $\mu(r)$. Overhead dash $(')$ denotes the derivative w.r.t radial co-ordinate $r$. To solve the eqs. (\ref{eq04})-(\ref{eq07}), we define the metric for interior space-time predicted by Finch-Skea \cite{FS} given below,
\begin{equation}
e^{2\mu(r)}=(1+Cr^2),\label{eq08}
\end{equation}

\begin{equation}
e^{2\nu(r)}=[(F+S \sqrt{1+Cr^2})Sin[\sqrt{1+Cr^2}]\label{eq09}
\end{equation}
$$+(S-F \sqrt{1+Cr^2})Cos[\sqrt{1+Cr^2}]^{2}.$$ 
where C, F and S are three arbitrary constants prescribing the specific geometry for 3-space of the interior space-time of the star and to be determined using the known boundary conditions for a stellar configuration. For a physically viable stellar model constants $C$, $F$ and $S$ must be real. The physical parameters namely energy density ($\rho$), radial $(p_{r})$, transverse pressure $(p_{t})$ and anisotropy $(\Delta)$ inside a compact star relevant to this model in D (i.e. n+2) dimensions are given by the following expressions:
\begin{equation}
\rho=\frac{Cn[n+1+(n-1)Cr^2]}{16\pi G_{D}y^2},\label{eq10}
\end{equation}

\begin{equation}
p_{r}=\frac{-Cn(nP+Q)}{16\pi G_{D}Py},\label{eq11}
\end{equation}

\begin{equation}
\Delta=\frac{(n-2)C^2r^2}{8\pi G_{D}y^2},\label{eq12}
\end{equation}
and
\begin{equation}
p_t = p_r + \Delta, \label{eq12a}
\end{equation}
where $y=1+Cr^2$, $P=(F+S\sqrt{y})~Sin\sqrt{y}+(S-F\sqrt{y})~Cos\sqrt{y}$ and $Q=(F\sqrt{y}-3S)~Cos\sqrt{y}-(3F+S\sqrt{y})~Sin\sqrt{y}$. Eqs. (\ref{eq08})-(\ref{eq12a}) are relevant to study the physical properties of a compact star. To obtain various physical quantities related to the compact objects in this model, we impose some boundary conditions given below,

\textbf{\bf{(a)}} At the surface ($r=b$) of a star one should match the interior solutions with the exterior metric given by Schwarzschild. In higher dimensions, the Schwarzschild exterior metric \cite{Shen} is given by, 
\begin{equation}
ds^2=-(1-\frac{W}{r^{n-1}})dt^{2}+(1-\frac{W}{r^{n-1}})^{-1}dr^{2}+r^2d\Omega_{n}^{2},\label{eq13}
\end{equation}
where $d\Omega_{n}^{2}=(d\theta_{1}^{2}+sin^{2}\theta_{1}~d\theta_{2}^{2}+sin^{2}\theta_{1}sin^{2}\theta_{2}d\theta_{3}^{2}+.....+sin^{2}\theta_{1}sin^{2}\theta_{2}sin^{2}\theta_{3}...sin^{2}\theta_{n-1}d\theta_{n}^{2}$) is defined earlier. In case of higher dimensional metric element, the constant $W$ appeared in eq. \ref{eq13} is connected to the mass $M$ of a star through the relation \cite{BCP} given below,

\begin{equation}
M=\frac{n A_{n}W}{16\pi G_{D}},\label{eq14}
\end{equation}
where $A_{n}=\frac{2 \pi^{\frac{n+1}{2}}}{\Gamma[\frac{n+1}{2}]}$. Therefore, if we consider the radius of a star is $b$, then from eqs. \ref{eq02} and \ref{eq13} one must have,
\begin{equation}
e^{2\nu(r=b)}=e^{-2\mu(r=b)}=(1-\frac{W}{b^{n-1}}).\label{eq15}
\end{equation}

\textbf{\bf{(b)}} The surface of a star is defined as the radial distance fron the centre of the star at which radial pressure ($p_r$) drops to zero. i.e at $r=b$, $p_{b}=0$. Therefore, from eq.~(\ref{eq11}), we get the following relation,
\begin{equation}
\frac{P(r=b)}{Q(r=b)}=-\frac{1}{n},\label{eq16}
\end{equation}

\textbf{\bf{(c)}} Here we choose the value of bag constant $B$ in the range $57.55 MeV/fm^3 < B < 95.11 MeV/fm^3$ or equivalently $145 MeV < B^{\frac{1}{4}} <164.4 MeV$ and the corresponding value of surface energy density ($\rho_s=4B$) is used to evaluate an upper limit on the radius of a star that may exist for which stable strange matter may be realized in four and higher dimensions. To get this bound, first we evaluated the necessary condition so that the constant $C$ in eq.~(\ref{eq08}) is real. This makes metric potential $\mu$ real. The value of $C$ can be calculated from eq.~(\ref{eq10}) for a given value of $B$ or equivalently surface energy density ($\rho_s$) and radius ($b$) of a compact object. From eq.~(\ref{eq10}), we note the value of constant $C$ and is given below,

\begin{equation}
C=\frac{(\delta-2b^2) \pm \sqrt{(\delta -2b^2)^2+2b^2(\gamma-2b^2)}}{ b^2(\gamma-2b^2)}.\label{eq17}
\end{equation}
The quantity $(\delta -2b^2)^2+2b^2(\gamma-2b^2)$ must be greater than or equal to zero so that the constant $C$ in eq.~(\ref{eq17}) is real. Here $\delta=\frac{1.5 \times 10^4 \times n(n+1)}{\rho_{s}}$ and $\gamma=\frac{3\times10^4 \times n(n-1)}{\rho_{s}}$ are two constants depends on space-time dimensions ($D=n+2$) and bag constant $B$ or surface energy density $\rho_{s}$. Therefore, the necessary and sufficient condition that the metric function $\mu$ to be real can be written as,

\begin{equation}
b\leq \frac{\delta}{\sqrt{2}}\sqrt{\frac{1}{2\delta+\gamma}}.\label{eq18}
\end{equation}

\textbf{\bf{(d)}} The causality condition must hold at all interior points and also at the surface of the star for a physically viable stellar model. This condition defines that the square of the radial sound velocity ($v_{r}^2$) and tangential sound velocity ($v_{t}^2$) throughout the interior fluid should be less than or equal to the velocity of light and should lie within the limit $0\leq v_{r}^2 \leq 1$ and $0\leq v_{t}^2 \leq 1$. This is alternatively equivalent to satisfy the following relations $0\leq\frac{dp_{r}}{d\rho}\leq1$ and $0\leq\frac{dp_{t}}{d\rho}\leq1$. 

\section{Numerical Analysis}
To calculate the maximum radius ($b_{max}$) and corresponding maximum mass of a strange star numerically, we discuss the techniques in this section. Here we consider the EOS of the interior matter content as MIT Bag model EOS having the value of $B$ within the range $57.55 MeV/fm^{3} < B < 95.11 MeV/fm^{3} $ necessary for strange matter which may be stable at zero external pressure relative to neutron \cite{Madsen}.

\begin{enumerate}
\item First, we calculate the radius of a star for a given value of surface energy density ($\rho_{s}$) through the relation $\rho_{s}=4B$ and space-time dimensions ($D$) using Eq.(\ref{eq18}) to get real value of $C$. In this calculation, we note that there exists two different values of $b$. In four dimensions ($D=4$), we note that out of these values of $b$, though we get real $C$, one value of $b$ and corresponding value of $C$ gives negative pressure as shown in Fig.(\ref{fig1}). So this value of $b$ is rejected as it gives non-physical result. For a physically viable stellar model, the following two criterion must hold. \textbf{(i)} pressure is positive and $(\frac{dp_r}{dr})\leq0$  and \textbf{(ii)} $(\frac{dp_{r}}{d\rho})\leq 1$ throughout the interior of the star. Other value of $b$ and $C$ satisfy the first condition but makes $(\frac{dp_{r}}{d\rho})> 1$. So the maximum allowed radius must be lower than this value of $b$ so that the second condition $(\frac{dp_{r}}{d\rho})\leq 1$ also hold when surface density ($\rho_{s}$) or Bag constant $B$ and space-time dimensions ($D$) are given. However, in higher dimensions greater than four $(D>4)$, we note that $p_r$ vary according to the physical condition but makes $(\frac{dp_{r}}{d\rho})> 1$ for one value of $b$. So it is also rejected.    

\begin{figure}
\centering
\includegraphics[width=8.3cm]{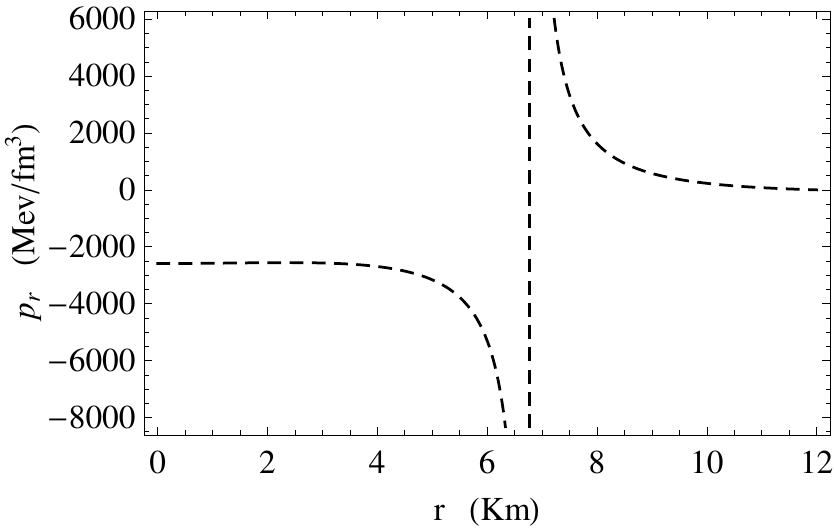}
\caption{$p_{r}$ Vs $r$ plot for $b=12$~km, $B=57.55$ $MeV/fm^{3}$ and $D=4$.}
\label{fig1}
\end{figure}

\begin{figure}
\centering
\includegraphics[width=8.3cm]{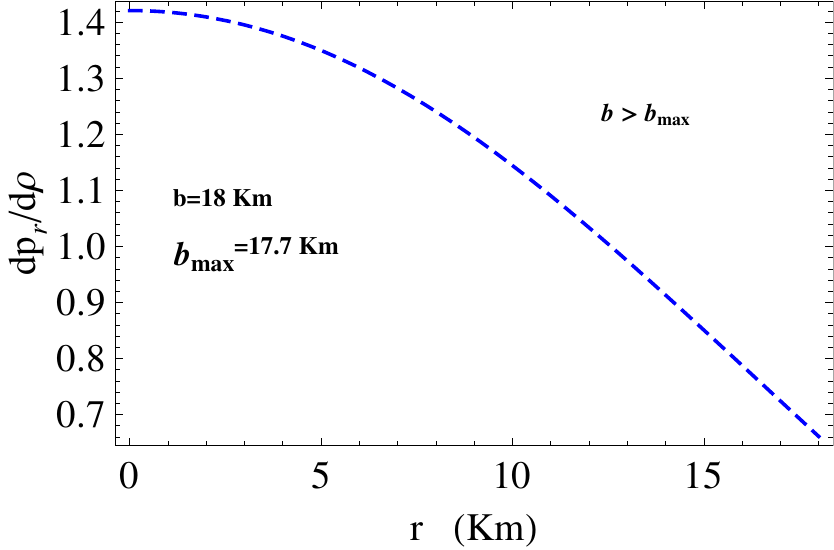}
\caption{$(\frac{dp_{r}}{d\rho})$ Vs $r$ plot for $b=18$~km, $B=57.55$ $MeV/fm^{3}$ and $D=5$.}
\label{fig2}
\end{figure}

\begin{table}
\centering
\caption{Values of maximum allowed radius ($b_{max}$), constant $C (km^{-2})$ and $y_{max}$ in different dimensions $(D)$ of a compact object for $B=57.55~MeV/fm^{3}$}\label{tab1}
\resizebox{.8\textwidth}{!}{$\begin{tabular}{@{}cccc}  \hline 
{Dimensions (D)}      & {$b_{max}~(km)$}         & {$C~(km^{-2})$} & {$y_{max}$} \\ \hline  
4     & $12.10$       & $0.017361$               & $2.54062$ \\ \cline{1-4} 
5     & $17.70$       & $0.003969$               & $1.24352$ \\ \cline{1-4} 
6     & $20.96$       & $0.001518$               & $0.67120$ \\ \cline{1-4} 
\end{tabular}$}
\end{table}

\begin{table}
\centering
\caption{Values of maximum allowed radius ($b_{max}$), constant $C (km^{-2})$ and $y_{max}$ in different dimensions $(D)$ of a compact object for $B=95.11~MeV/fm^3$}\label{tab2}
\resizebox{.8\textwidth}{!}{$\begin{tabular}{@{}cccc}  \hline
{Dimensions (D)}     & {$b_{max}~(km)$}      & {$C (km^{-2})$}    & {$y_{max}$} \\ \hline  
4                    & $9.41$                &$0.028705$          & $2.5418$ \\ \cline{1-4} 
5                    & $13.77$               &$0.006561$          & $1.2439$ \\ \cline{1-4} 
6                    & $16.31$               &$0.002525$          & $0.6713$ \\ \cline{1-4} 
            
\end{tabular}$}
\end{table}

\begin{figure}
\centering
\includegraphics[width=8.3cm]{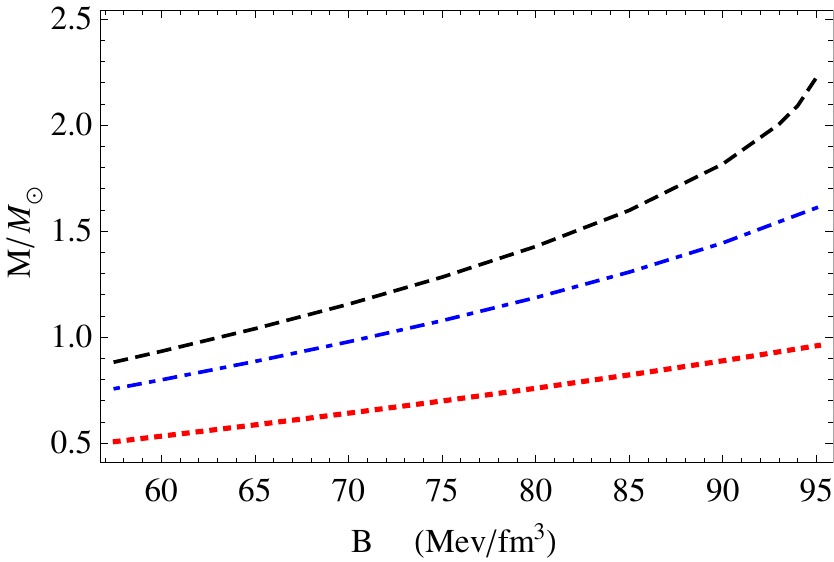}
\caption{Variation of Mass $M~(M_{\odot})$ with bag constant $(B)$ for $D=4$. Lines from top to bottom represent respectively for b = 9.4, 9, 8~(km).}
\label{fig3}
\end{figure}

\begin{figure}
\centering
\includegraphics[width=8.3cm]{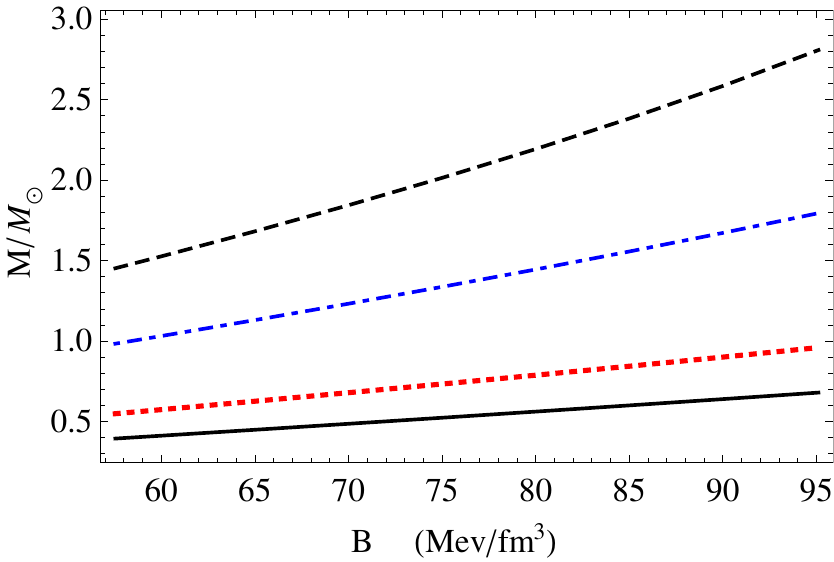}
\caption{Variation of Mass $M~(M_{\odot})$ with Bag constant $B$ for $D=5$. Lines from top to bottom represent respectively for b = 13.5, 12, 10, 9~(km).}
\label{fig4}
\end{figure}

\begin{figure}
\centering
\includegraphics[width=8.3cm]{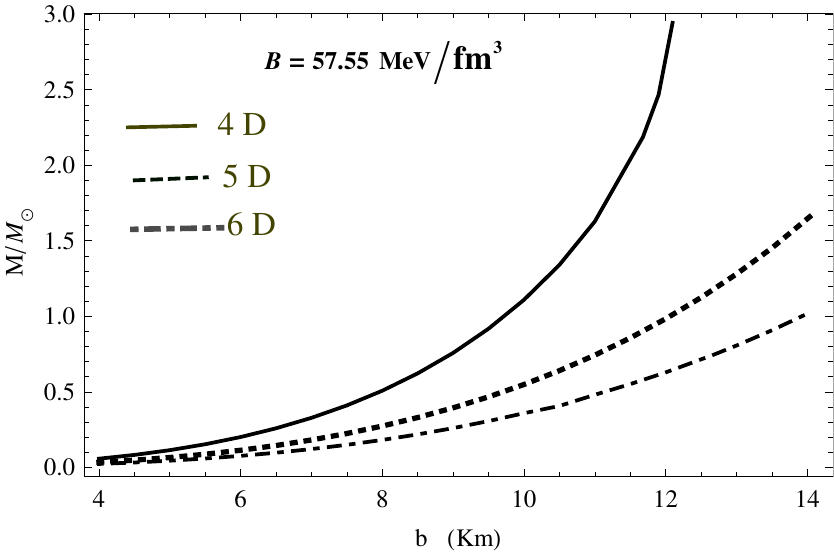}
\caption{Mass-radius plot for $B=57.55~MeV/fm^3$. Solid, dotted and dot dashed line represent respectively for $D=4$, $5$ and $6$.}
\label{fig5}
\end{figure}

\begin{figure}
\centering
\includegraphics[width=8.3cm]{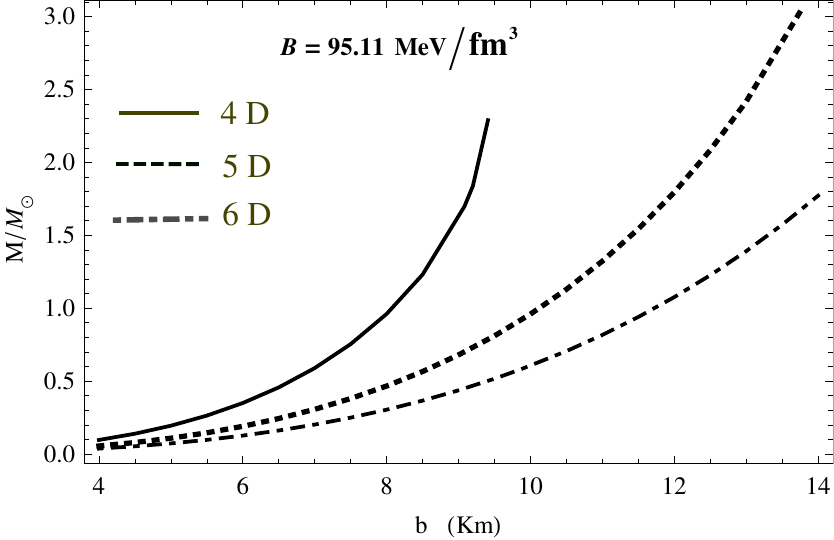}
\caption{Mass-radius for $B=95.11~MeV/fm^3$. Solid, dotted and dot dashed line for $D=4$, $5$ and $6$ respectively.}
\label{fig6}
\end{figure}

\item To evaluate the maximum allowed radius in the present model ($b_{max}$) that makes $p_{r}$ to vary according to the physical requirements of a star as well as $(\frac{dp_{r}}{d\rho})\leq1$ throughout the star, we consider $(\frac{dp_{r}}{d\rho})_{max}=1$ at $r=0$ of the star. This gives the value of maximum allowed radius $b_{max}$ for a given surface density $(\rho_{s}=4B)$ and space-time dimensions $(D)$ and we evaluated the corresponding value of $C$. We have checked that when $b > b_{max}$, $(\frac{dp_{r}}{d\rho})>1$ which is shown in figure (\ref{fig2}). The maximum mass $(M_{max})$ (in unit of $M_{\odot}$) of a star is calculated within the radius $b_{max}$ for a given $\rho_{s}$ or $B$ and $D$ in this model.
\item We choose the range of $B$ as $57.55~MeV/fm^{3}\leq B\leq 95.11~MeV/fm^{3}$ to study the variations of mass-radius relationship of of strange star and their other relevant physical properties.
\end{enumerate}

\section{Mass-radius Relationship}
The total mass of a compact star contained within a radius $b$ in D-dimensional space-time is given by,
\begin{equation}
M=A_{n}\int^{b}_{0}r^{D-2} \rho(r)dr,\label{eq19}
\end{equation}
where $A_{n}=\frac{2\pi^{\frac{n+1}{2}}}{\Gamma(\frac{n+1}{2})}$ and expression for energy density $\rho(r)$ is given by eq. (\ref{eq10}). From eq. (\ref{eq10}) and (\ref{eq19}), we note that in this model the total mass $(M)$ depends on the constant $C$ defined in eq. (\ref{eq08}) and on the space-time dimensions $D$. We have determined the total mass of a star contained within a radius $b$ for different values of surface density or $B$ through the relation $\rho_{s}=4B$ in four and higher dimensions. First, we have picked up a value of $B$ within the range of $57.55~MeV/fm^3<B<95.11~MeV/fm^3$ necessary for stable strange matter relative to neutron at zero pressure (at the surface) condition \cite{Madsen} and corresponding value of $\rho$ is taken to evaluate the value of maximum allowed radius $b_{max}$ for a given space-time dimensions $(D)$. Maximum allowed radii $b_{max}$ are tabulated in tables (\ref{tab1}) and (\ref{tab2}).  From tables (\ref{tab1}) and (\ref{tab2}), we note that in four dimensions the maximum alowed radius of a strange star may varies between $9.41-12.10~km$ when $B$ varies from $95.11-57.55~MeV/fm^3$ respectively. However, maximum radius increases with the increase of space-time dimensions. Now constant $C$ can be evaluated from eq. (\ref{eq10}). As constant $C$ and radius b ($<b_{max}$) are known, we can evaluate the mass of a compact star using eqs. (\ref{eq14}) and (\ref{eq15}) for known value of $B$ and $D$. Keeping radius $(b)$ fixed at different values ($\leq b_{max}$), we have studied the variation of mass of a star with bag parameter $(B)$ in different dimensions $(D)$ and the results are shown in figures (\ref{fig3}) and (\ref{fig4}). In this paper, we have also studied the mass-radius relation of a star for $B=57.55~MeV/fm^3$ and $95.11~MeV/fm^3$ and are plotted in the figures (\ref{fig5}) and (\ref{fig6}) respectively for three different dimensions $D=4$, $5$ and $6$. From figures (\ref{fig3}) and (\ref{fig4}), we note that the mass of a star $M$ increases with increase of bag constant ($B$) when radius $(b)$ and space-time dimensions $(D)$ are fixed at some values. 

\subsection{Maximum mass and Surface red-shift}
To determine maximum mass of a strange star of radius $b$ in $D$-dimensional space-time in this model, we follow the approach derived by Paul \cite{BCP}. In the D-dimensional space-time is the mass of a compact star contained within the radius $b$ is given by:  

\begin{equation}
M(b)=\frac{n A_{n} C b^{n+1}}{16\pi G_{D}(1+Cb^2)}.\label{eq20}
\end{equation}
This can be written as in the form 

\begin{equation}
M(b)=\frac{n A_{n} y b^{n-1}}{16\pi G_{D}(1+y)},\label{eq21}
\end{equation}
where $y=Cb^2$ and the corresponding compactness is given by 
\begin{equation}
u = \frac{M(b)}{b} =\frac{n A_{n} b^{n-2}}{16\pi G_{D}(\frac{1}{y}+1)}.\label{eq22} 
\end{equation}
From eqs. (\ref{eq21}) and (\ref{eq22}), we note that the when $b=b_{max}$, both the mass ($M$) and compactness $(u)$ are maximum for which y $(=C~b^2)$ is also maximum. Therefore maximum compactness ($u_{max}$) and hence maximum mass ($M_{max}$) is given by

\begin{equation}
u_{max} = \frac{M_{max}}{b_{max}}=\frac{n A_{n} b_{max}^{n-2}}{16\pi G_{D}(\frac{1}{y_{max}}+1)}.\label{eq22a} 
\end{equation}

\begin{table}
\centering
\caption{Maximum radius ($b_{max}$) and maximum mass $M_{max}(M_{\odot})$, compactness($u_{max}$)  and maximum surface red-shift $(Z_{s})_{max}$ for $B=57.55~MeV/fm^3$ satisfying causality conditions.}\label{tab3}
\resizebox{.8\textwidth}{!}{$
\begin{tabular}{@{}ccccc}  \hline
{Dimensions(D)} & {$b_{max}~(km)$} & {$M_{max}~(M_{\odot})$} & {$u_{max}$} &{$(Z_{s})_{max} $} \\ \hline 
4     & $12.10$                &$2.94$                   & $0.35878$    & $0.8822$  \\ \cline{1-5} 
5     & $17.70$                &$3.92$                   & $0.32649$    & $0.6976$  \\ \cline{1-5} 
6     & $20.96$                &$3.85$                   & $0.26775$    & $0.4673$  \\ \cline{1-5} 
\end{tabular}$}
\end{table}

\begin{table}
\centering
\caption{Maximum radius ($b_{max}$) and maximum mass $M_{max}(M_{\odot})$, compactness ($u_{max}$) and maximum surface red-shift $(Z_{s})_{max}$ for $B=95.11~MeV/fm^3$ satisfying causality condition.}\label{tab4}
\resizebox{.8\textwidth}{!}{$
\begin{tabular}{@{}ccccc}  \hline
{Dimensions (D)} & {$b_{max}~(km)$} & {$M_{max}~(/M_{\odot})$} & {$u_{max}$} & {$(Z_{s})_{max} $} \\  \hline 
4     & $9.41$          & $2.29 $         & $0.35883$     & $0.8819$   \\ \cline{1-5} 
5     & $13.77$         & $3.05 $         & $0.32654$     & $0.6978$   \\ \cline{1-5} 
6     & $16.30$         & $2.96 $         & $0.26778$     & $0.4674$   \\ \cline{1-5} 
\end{tabular}$}
\end{table}

\begin{figure}
\centering
\includegraphics[width=8.3cm]{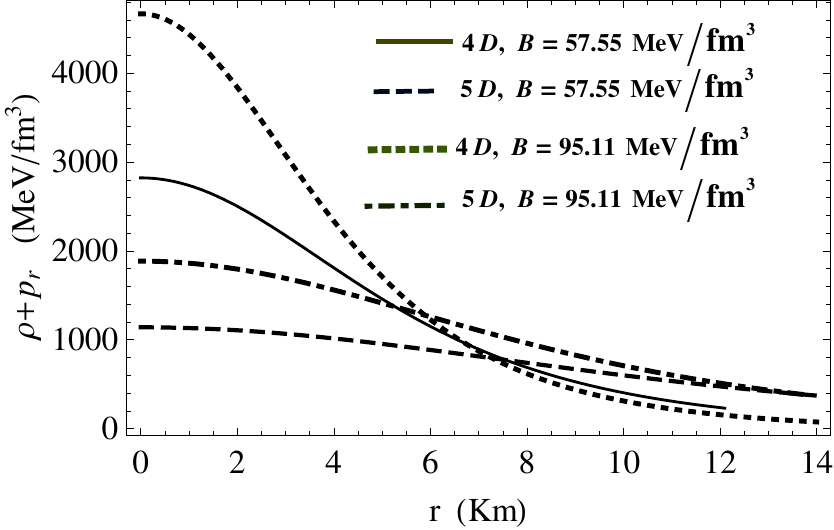}
\caption{Plot of $(\rho +p_{r})$ with  $r$. Solid line for $D=4$, $B=57.55~MeV/fm^3$, dotted line for $D=4$, $B=95.11~MeV/fm^3$ and dashed line for $D=5$, $B=57.55~MeV/fm^3$, dot dashed line for $D=5$, $B=95.11~MeV/fm^3$.}
\label{fig7}
\end{figure}

\begin{figure}
\centering
\includegraphics[width=8.3cm]{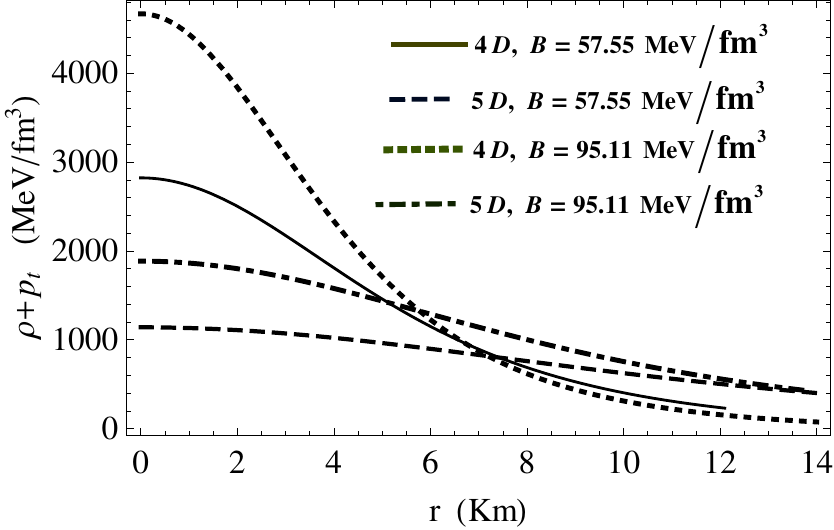}
\caption{Plot of $(\rho +p_{t})$ with  $r$. Solid line for $D=4$, $B=57.55~MeV/fm^3$, dotted line for $D=4$, $B=95.11~MeV/fm^3$ and dashed line for $D=5$, $B=57.55~MeV/fm^3$, dot dashed line for $D=5$, $B=95.11~MeV/fm^3$.}
\label{fig8}
\end{figure}

\begin{figure}
\centering
\includegraphics[width=8.3cm]{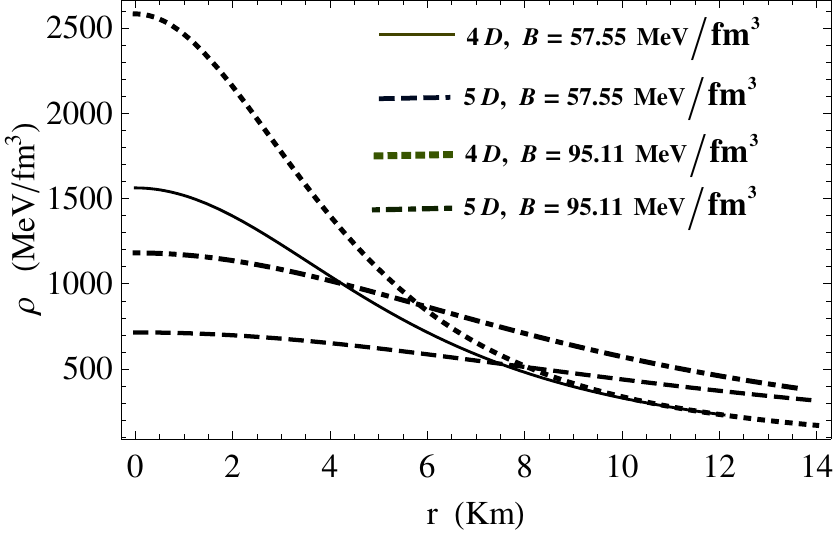}
\caption{Plot of energy density $(\rho)$ with  $r$. Solid line for $D=4$, $B=57.55~MeV/fm^3$, dotted line for $D=4$, $B=95.11~MeV/fm^3$ and dashed line for $D=5$, $B=57.55~MeV/fm^3$, dot dashed line for $D=5$, $B=95.11~MeV/fm^3$.}
\label{fig9}
\end{figure}

\begin{figure}
\centering
\includegraphics[width=8.3cm]{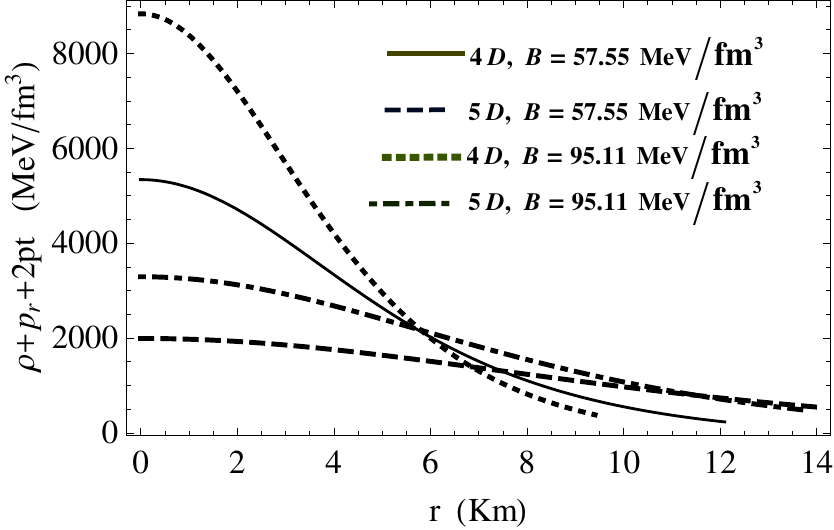}
\caption{Plot of $(\rho +p_{r}+2p_{t})$ with  $r$. Solid line for $D=4$, $B=57.55~MeV/fm^3$, dotted line for $D=4$, $B=95.11~MeV/fm^3$  and dashed line for $D=5$, $B=57.55~MeV/fm^3$, dot dashed line for $D=5$, $B=95.11~MeV/fm^3$.}
\label{fig10}
\end{figure}

\begin{figure}
\centering
\includegraphics[width=8.3cm]{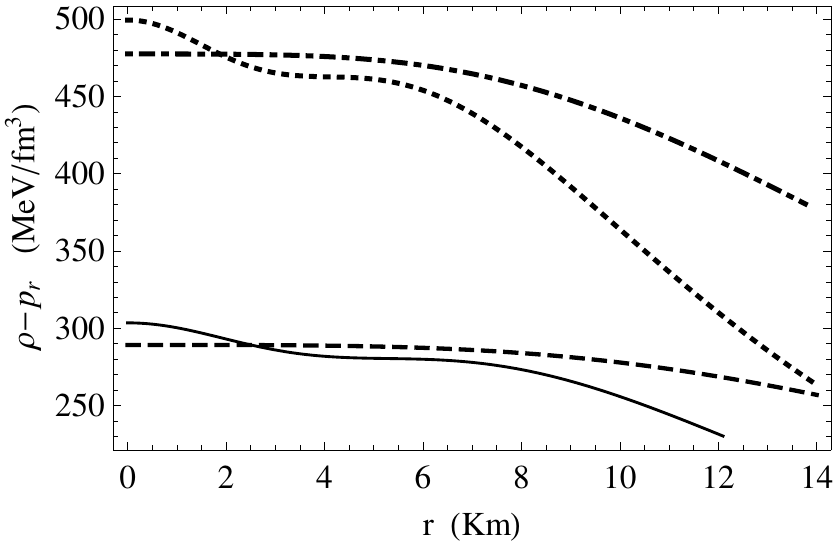}
\caption{Plot of $(\rho -p_{r})$ with  $r$. Solid line for $D=4$, $B=57.55~MeV/fm^3$, dotted line for $D=4$, $B=95.11~MeV/fm^3$ and dashed line for $D=5$, $B=57.55~MeV/fm^3$, dot dashed line for $D=5$, $B=95.11~MeV/fm^3$.}
\label{fig11}
\end{figure}

The maximum surface redshift ($Z_{s}$) can be obtained following the work of Paul et al \cite{BCPAUL} and is given by,
\begin{equation}
(Z_{s})_{max}=(1-2u_{max})^{-\frac{1}{2}}-1\label{eq22b} 
\end{equation}

We first evaluated the maximum value of the radius ($b_{max}$) by using the method discussed in \S 3. Accordingly, we note the maximum value of the parameter $y$ defined by $y = C~b_{max}^2$. The corresponding maximum mass ($M_{max}$) and compactness ($u_{max}$) are calculated using eq. (\ref{eq22a}). We have determined the values of $b_{max}$, $C$ and $y_{max}$ and are tabulated in tables (\ref{tab1}) and (\ref{tab2}) for $B=57.55$ and $95.11~MeV/fm^3$ in different space-time dimensions ($D$). We have studied the effect of dimensions $D$ on the value of maximum radius $(b_{max})$, corresponding maximum mass ($M_{max}$), compactness ($u_{max}$) and surface red-shift $(Z_{s})$ for $B=57.55$ and $95.11~MeV/fm^3$ and are tabulated in tables (\ref{tab3}) and (\ref{tab4}). Buchdahl \cite{Buchdahl} proposed analytically that maximum compactness (mass to radius ratio) of a compact object in four dimension should be $\frac{M}{b}\leq \frac{4}{9}$. Ponce De Leon and Cruz \cite{PDL} extends the Buchdahl limit in $D$ dimensional space-time which is $\frac{M}{b^{(D-3)}}\leq \frac{2(D-2)}{(D-1)^2}$. From tables (\ref{tab3}) and (\ref{tab4}), we note that maximum compactness $(u_{max})$ obtained in this model always lies below the allowed limit \cite{Buchdahl,PDL}.

\subsection{Energy Conditions}
\begin{figure}
\centering
\includegraphics[width=8.3cm]{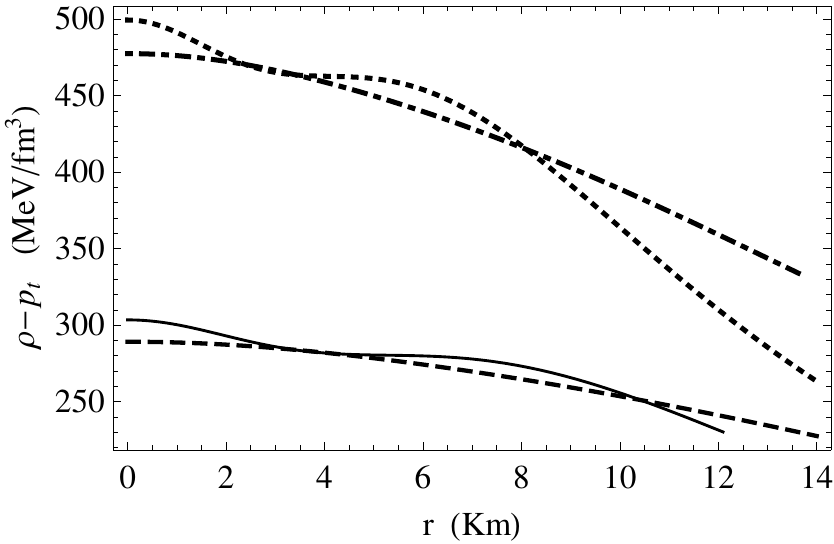}
\caption{Plot of $(\rho-p_{t})$ with $r$. Solid line for $D=4$, $B=57.55~MeV/fm^3$, dotted line for $D=4$, $B=95.11~MeV/fm^3$ and dashed line for $D=5$, $B=57.55~MeV/fm^3$, dot dashed line for $D=5$, $B=95.11~MeV/fm^3$.}
\label{fig12}
\end{figure}

For a physically viable stellar model the following energy conditions \cite{Carrol,Pant} should be satisfied from the centre upto the surface $(b_{max})$ of a star for chosen configurations. We have evaluated the following energy conditions numerically in this model:
\begin{enumerate}
\item Null Energy Condition (NEC: $\rho + p_{r}  \geq 0$, $\rho + p_{t} \geq 0$)
\item Weak Energy Condition (WEC: $\rho + p_{r} \geq 0$, $\rho \geq 0$, $\rho + p_{t} \geq 0$ )
\item Strong Energy Condition (SEC: $\rho + p_{r} \geq 0$, $\rho +  p_{r}+ 2p_{t} \geq 0$)
\item Dominant Energy Condition (DEC: $\rho \geq 0$, $\rho-p_{r} \geq 0$,  $\rho-p_{t} \geq 0$)
\end{enumerate}
Null Energy Conditions (NEC) are shown in figures (\ref{fig7}) and (\ref{fig8}). Weak Energy Conditions (WEC) are plotted in figures  (\ref{fig7}), (\ref{fig8}) and (\ref{fig9}). Strong Energy Conditions (SEC) are plotted in figures (\ref{fig7}) and (\ref{fig10}). Dominant Energy Conditions (DEC) are plotted in figures (\ref{fig9}), (\ref{fig11}) and (\ref{fig12}). From figures (\ref{fig7})-(\ref{fig12}), we found that all necessary energy conditions are satisfied in the present model for the bag parameter $B=57.55$ and $B=95.11~MeV/fm^{3}$ in four and highr dimensions.
 
\begin{figure}
\centering
\includegraphics[width=8.3cm]{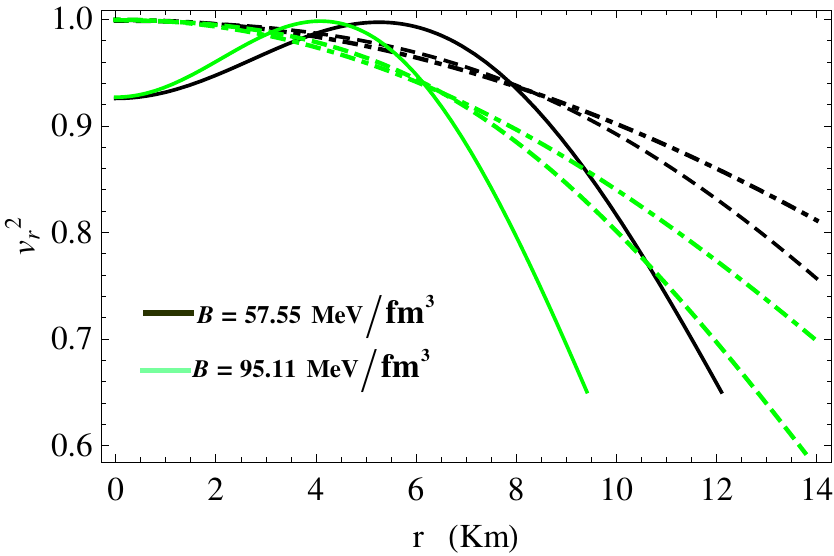}
\caption{Plot of $(v^2_{r})$ with $r$. Black and green colours represent the plots for $B=57.55$ and $95.11~MeV/fm^3$ respectively. Solid, dashed and dot dashed line represent respectively for $D=4$, $D=5$ and $D=6$.}
\label{fig13}
\end{figure}

\begin{figure}
\centering\includegraphics[width=8.3cm]{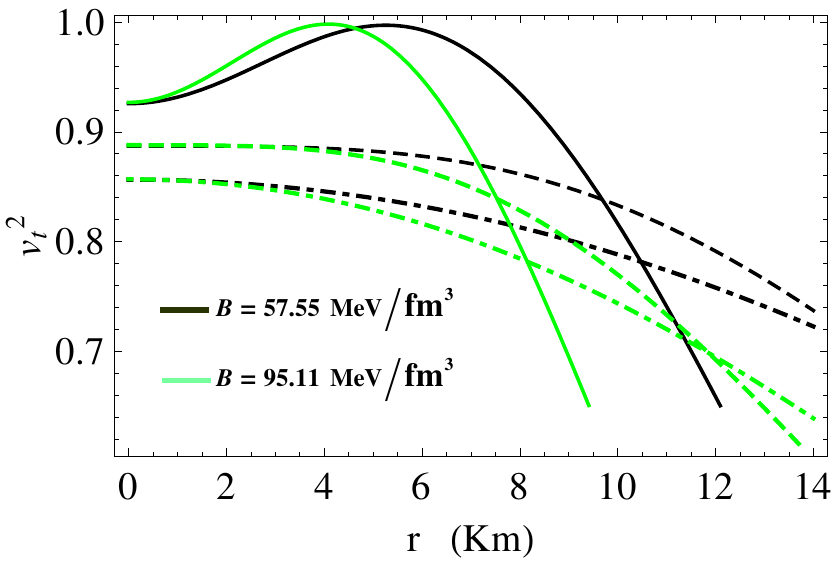}
\caption{Plot of $(v^2_{t})$ with $r$. Black and green colours represent the plots for $B=57.55$ and $95.11~MeV/fm^3$ respectively. Solid, dashed and dot dashed line represent respectively for $D=4$, $D=5$ and $D=6$.}
\label{fig14}
\end{figure}

\begin{table}
\centering
\caption{Values of parameters $C$, $F$ and $S$ for VELA X1 for different $(B)$ and dimensions (D).}\label{tab6}
\resizebox{.8\textwidth}{!}{$
\begin{tabular}{@{}ccccc}  \hline
{Dimensions(D)} & {$B~(MeV/fm^3)$}    & {$C~(km^{-2})$}    & {$F$}          & {$S$}          \\  \hline 
4               & 85.51               & $0.0131746$        & $0.316051$      & $0.255032$    \\ \cline{1-5} 
5               & 175.4               & $0.0094338$        & $0.717021$      & $0.149222$    \\ \cline{1-5} 
6               & 281.1               & $0.0075923$        & $1.189520$      & $0.022099$    \\ \cline{1-5} 
\end{tabular}$}
\end{table}

\begin{table}
\centering
\caption{The mass $M(M_{\odot})$ and compactness ($u$) of VELA X1 for different $(B)$ and dimensions (D).}\label{tab5}
\resizebox{.8\textwidth}{!}{$ 
\begin{tabular}{@{}ccccc}  \hline
{Dimensions(D)}   & {$B~(MeV/fm^3)$}  & {$b_{max}~(km)$}    & {$M~(M_{\odot})$}  & {$u$}         \\  \hline 
4                 &    85.51          & $9.56$              &$1.77$              & $0.2731$      \\ \cline{1-5} 
5                 &    175.4          & $9.56$              &$1.77$              & $0.2731$      \\ \cline{1-5} 
6                 &    281.1          & $9.56$              &$1.77$              & $0.2731$      \\ \cline{1-5} 
\end{tabular}$}
\end{table}

\subsection{Causality Condition}
We have check the validity of causality conditions for the chosen range of values of $B~(57.55-95.11~MeV/fm^3)$ required for stable strange matter relative to neutron at zero external pressure \cite{Madsen}. We have determined the numerical value of $v_{r}^2$ and $v_{t}^2$ from centre to surface of the star for three different values of space-time dimensions $D=4$, $5$ and $6$ and are shown in figures (\ref{fig13}) and (\ref{fig14}). From figures (\ref{fig13}) and (\ref{fig14}), we note that causality conditions are satisfied throughout the star. 

\begin{figure}
\centering
\includegraphics[width=8.3cm]{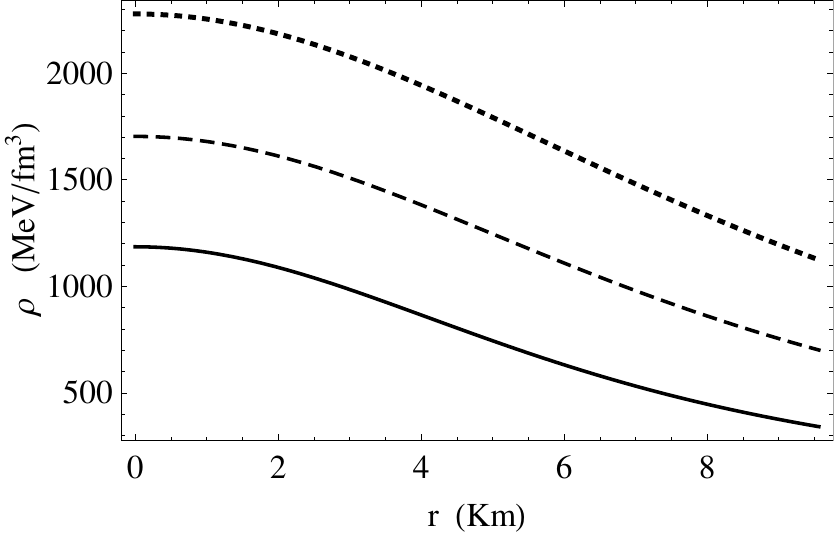}
\caption{Variations of energy density ($\rho$) with radial distance $(r)$ for VELA X-1. Solid line for $D=4$, $B=85.51~MeV/fm^3$. Dashed line for $D=5$, $B=175.4~MeV/fm^3$ and doted line for $D=6$, $B=281.1~MeV/fm^3$.}
\label{fig15}
\end{figure}

\begin{figure}
\centering
\includegraphics[width=8.3cm]{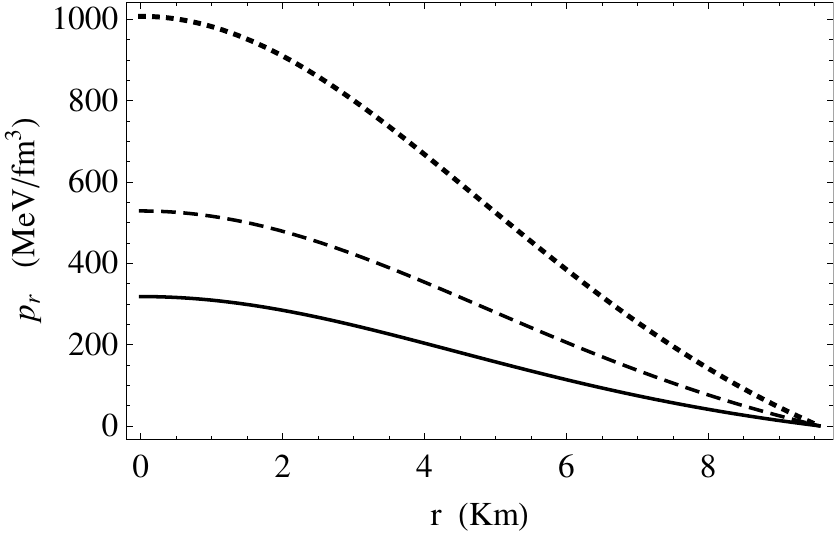}
\caption{Variations of $p_r$ with radial distance $(r)$ for VELA X-1. Solid line for $D=4$, $B=85.51~MeV/fm^3$. Dashed line for $D=5$, $B=175.4~MeV/fm^3$ and doted line for $D=6$, $B=281.1~MeV/fm^3$.}
\label{fig16}
\end{figure}

\begin{figure}
\centering
\includegraphics[width=8.3cm]{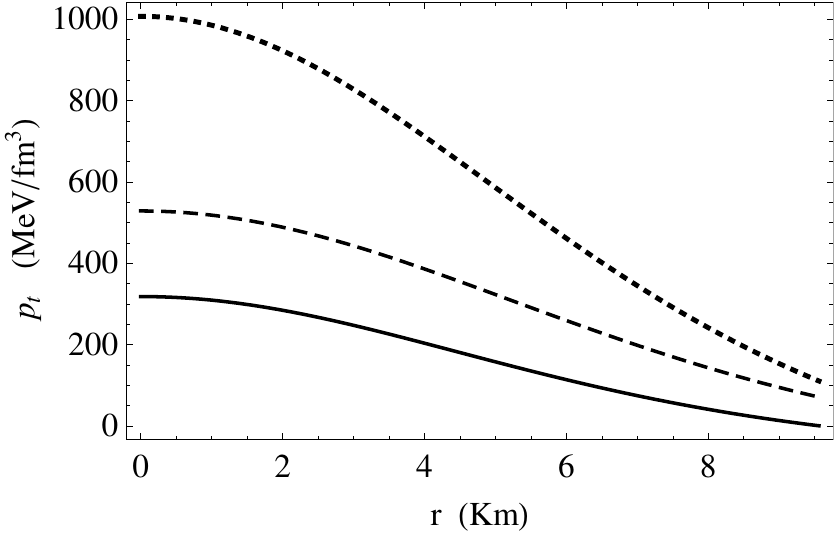}
\caption{Variations of $p_t$ with radial distance $(r)$ for VELA X-1. Solid line for $D=4$, $B=85.51~MeV/fm^3$. Dashed line for $D=5$, $B=175.4~MeV/fm^3$ and doted line for $D=6$, $B=281.1~MeV/fm^3$.}
\label{fig17}
\end{figure}

\begin{figure}
\centering
\includegraphics[width=8.3cm]{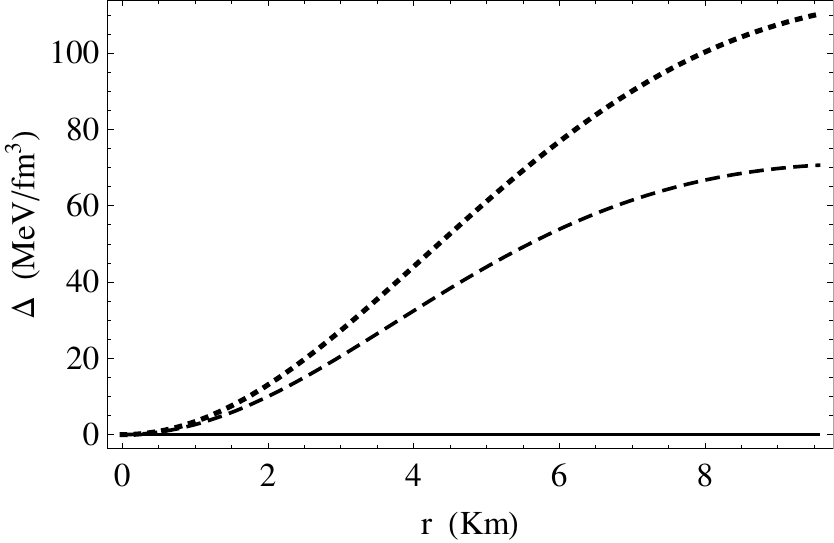}
\caption{Variations of Anisotropy ($\Delta$) with radial distance $(r)$ for VELA X-1. Solid line for $D=4$,$B=85.51~MeV/fm^3$. Dashed line for $D=5$, $B=175.4~MeV/fm^3$ and doted line for $D=6$, $B=281.1~MeV/fm^3$.}
\label{fig18}
\end{figure}

\section{Physical Applications}
Let us now apply this model to study the physical behaviour of few compact objects which are supposed to be strange stars. We considered here two different compact objects of known radius ($b$) and mass ($M$). For the given mass and radius we can solve the eq. (\ref{eq15}) to evaluate the value of constant $C$ for metric function $\mu$. Simultaneously solving eqs. (\ref{eq15}) and (\ref{eq16}), we get the values of constants $F$ and $S$. Now we can study the variations of $\rho, p_{r}, p_{t}, \Delta$ and other parameters related to the compact stars. To study the effect of bag constant ($B$) on the mass ($M$) of a compact star, we adopt here a different approach in which the surface energy density $\rho_{s}$ of a compact object is assumed corresponding to the value of $B~(\rho_{s}=4B)$ within the range $57.55<B~(MeV/fm^3)<95.11$ necessary for stable bulk strange star. Now constant $C$ can be evaluated from eq. (\ref{eq10}). Subsequently we can obtain the mass($M$) of compact strange star from eq. (\ref{eq14}) and study the effect of bag constant ($B$) on mass ($M$). 
  
\textbf{Case I:} We consider here the compact star VELA X-1 \cite{Rawls} having observed mass $M=1.77~M_{\odot}$ and radius $b=9.56~km$. In table \ref{tab6}, we have tabulated the values of constants $C$, $F$ and $S$ for chosen combinations of bag constant ($B$) and space-time dimensions ($D$) satisfying all the physical conditions for VELA X-1 in this model. Using the values of constants given in table (\ref{tab6}), we have studied the variation of energy density ($\rho$), radial ($p_r$), transverse pressure ($p_t$) and anisotropy ($\Delta$) and are shown in figures (\ref{fig15}) - (\ref{fig18}) respectively. In this model one can predict the observed mass of a strange star like VELA X-1 by adjusting the value of bag constant $B$ and dimensions $D$ when radius of the star is taken at observed value. We have predicted the mass of VELA X-1 using eq. (\ref{eq14}) for observed radius and tabulated in the table \ref{tab5}. From table \ref{tab5} it is observed that the mass $M=1.77~M_{\odot}$ of VELA X-1 may be predicted for different combinations of space-time dimensions ($D$) and bag constant ($B$). Such as in four dimensions, the mass $M=1.77~M_{\odot}$ may be predicted with $B=85.51~MeV/fm^3$ for radius $b=9.56~km$. When $D=5$, observed mass $1.779M_{\odot}$ predicted with $B=175.4~MeV/fm^3$. Similarly for $D=6$, we note that the observed mass $1.77M_{\odot}$ may be realised for $B=281.1~MeV/fm^3$. 

\begin{table}
\centering
\caption{Values of parameters $C$, $F$ and $S$ of 4U 1538-52 for different values of bag constant $(B)$ and space-time dimensions (D).}\label{tab8}
\resizebox{.8\textwidth}{!}{$
\begin{tabular}{@{}ccccc}  \hline
{Dimensions(D)} & {$B~(MeV/fm^3)$} & {$C~(km^{-2})$}   & {$F$}         & {$S$}     \\  \hline 
4               &92.85             & $0.0078288$       &$0.26890$      &$0.45781$  \\ \cline{1-5} 
5               &173.54            & $0.0061911$       &$0.78492$      &$0.32702$  \\ \cline{1-5} 
6               &267.61            & $0.0052370$       &$1.34428$      &$0.18692$  \\ \cline{1-5} 
\end{tabular}$}
\end{table}

\begin{table}[h!]
\centering
\caption{The mass $M(M_{\odot})$ and compactness ($u$) of 4U 1538-52 for different values of bag constant $(B)$ and space-time dimensions (D).}\label{tab7}
\resizebox{.8\textwidth}{!}{$
\begin{tabular}{@{}ccccc}  \hline
{Dimensions(D)} & {$B~(MeV/fm^3)$} & {$b_{max}~(km)$} & {$M~(M_{\odot}$)} & {$u$}         \\  \hline
4               &  92.85         & $7.866$          &$0.87 $            & $0.1631$      \\ \cline{1-5} 
5               &  173.54        & $7.866$          &$0.87 $            & $0.1631$      \\ \cline{1-5} 
6               &  267.61        & $7.866$          &$0.87 $            & $0.1631$      \\ \cline{1-5} 
\end{tabular}$}
\end{table}

\begin{figure}
\centering
\includegraphics[width=8.3cm]{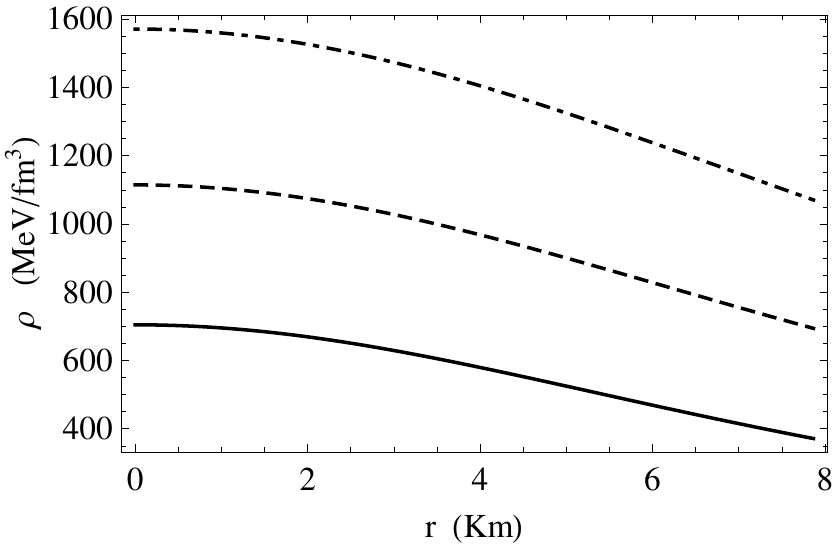}
\caption{Variations of energy density ($\rho$) with radial distance $(r)$ for 4U 1538-52. Solid line for $D=4$, $B=92.85~MeV/fm^3$. Dashed line for $D=5$, $B=173.58~MeV/fm^3$ and Dot Dashed line for $D=6$, $B=267.61~MeV/fm^3$.}
\label{fig19}
\end{figure}

\begin{figure}
\centering
\includegraphics[width=8.3cm]{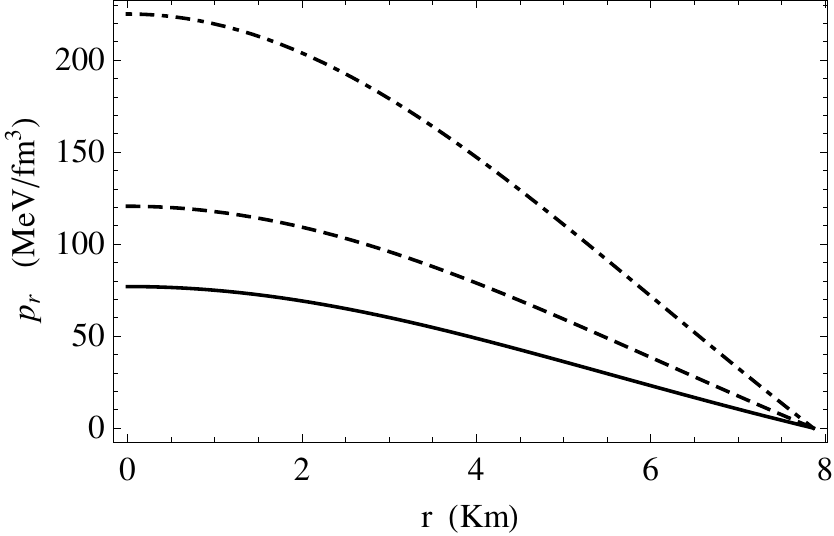}
\caption{Variations of $p_r$ with radial distance $(r)$ for 4U 1538-52. Solid line for $D=4$, $B=92.85~MeV/fm^3$. Dashed line for $D=5$, $B=173.58~MeV/fm^3$ and Dot Dashed line for $D=6$, $B=267.61~MeV/fm^3$.}
\label{fig20}
\end{figure}

\begin{figure}
\centering
\includegraphics[width=8.3cm]{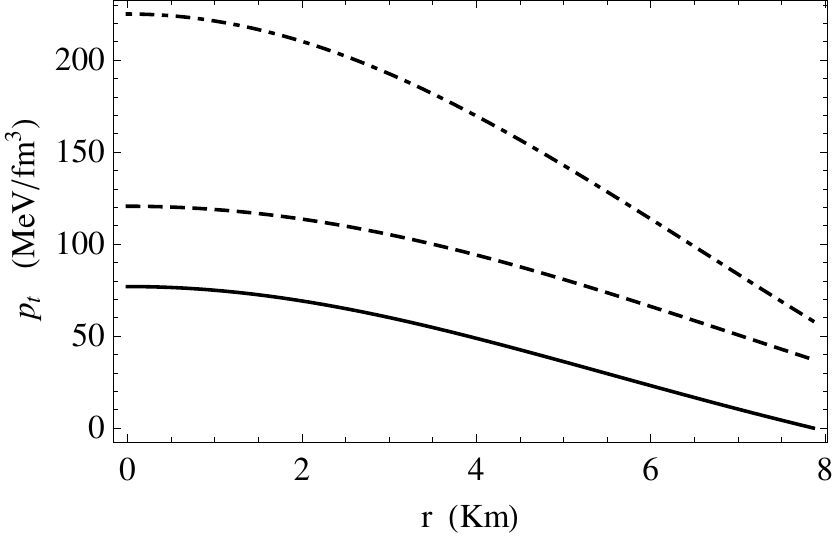}
\caption{Variations of $pt$ with radial distance $(r)$ for 4U 1538-52. Solid line for $D=4$, $B=92.85~MeV/fm^3$. Dashed line for $D=5$, $B=173.58~MeV/fm^3$ and Dot Dashed line for $D=6$, $B=267.61~MeV/fm^3$.}
\label{fig21}
\end{figure}

\begin{figure}
\centering
\includegraphics[width=8.3cm]{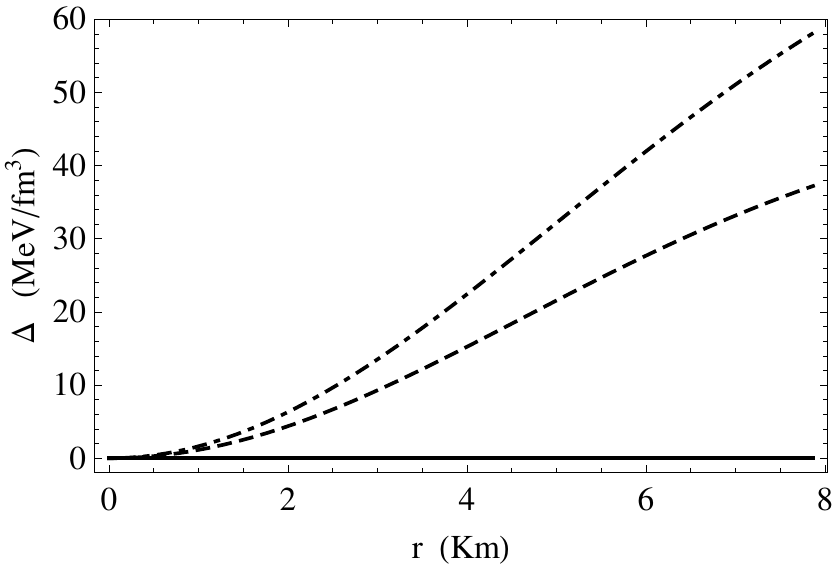}
\caption{Variations of Anisotropy ($\Delta$) with radial distance $(r)$ for 4U 1538-52. Solid line for $D=4$, $B=92.85~MeV/fm^3$. Dashed line for $D=5$, $B=173.58~MeV/fm^3$ and Dot Dashed line for $D=6$, $B=267.61~MeV/fm^3$.}
\label{fig22}
\end{figure}

\textbf{Case-II:} Next we consider the star 4U 1538-52 \cite{Rawls} having observed mass $M=0.87~M_{\odot}$ and radius $b=7.866~km$ which is also supposed to be a strange star. Using the values of $C$, $F$ and $S$ for 4U 1538-52 tabulated in table-\ref{tab8}, we have studied the variations of different physical parameters such as energy density $(\rho)$, radial $(p_r)$, transverse pressure $(p_t)$ and anisotropy $(\Delta)$ and are shown in figures (\ref{fig19}) - (\ref{fig22}) respectively. From figures (\ref{fig19})-(\ref{fig21}), we note that energy density ($\rho$) and pressures ($p_{r}$, $p_{t}$) increases with the increase of space-time dimensions ($D$) in this model. In figure(\ref{fig22}), it is evident that anisotropy ($\Delta$) picks up higher values for higher space-time dimensions ($D$). But in four dimensions ($D=4$) the star is found to be isotropic as $\Delta=0$ throughout the star. We have also predicted the mass of 4U 1538-52 using eq.(\ref{eq14}) for observed radius $b=7.866~km$. The mass ($M$) of 4U 1538-52 have been tabulated in table-\ref{tab7}. We note that the mass $0.87M_{\odot}$ may be predicted with the different choice of $D$ and $B$ as in case of VELA X-1. In $D=4$ we may get the mass $M=0.87M_{\odot}$ for radius $b=7.866$ Km and $B=92.85$ $MeV/fm^3$. When $D=5$ we observed the mass $M=0.87M_{\odot}$ for radius $b=7.866$ Km and $B=173.54$ $MeV/fm^3$. Similarly in six dimensions the mass $M=0.87M_{\odot}$ for radius $b=7.866$ Km may be predicted when $B=267.61$ $MeV/fm^3$. In table \ref{tab8} we have tabulated the values of parameters $C$,$F$ given by eq.(\ref{eq08}) and eq.(\ref{eq09}) for specific combinations of Bag parameter($B$) and space-time dimensions $(D)$ to satisfy all physical conditions for 4U 1538-52 in this model.

\section{Stability analysis of the model}
The stability analysis for the stellar configuration in this model from the following points of views are discussed in this section:
\begin{enumerate}
\item  Stability analysis using TOV equation 
\item  Stability analysis using Herrera cracking condition and 
\item  Stability analysis using Adiabatic index 
\end{enumerate}

\begin{figure}[h!]
\centering
\includegraphics[width=8.3cm]{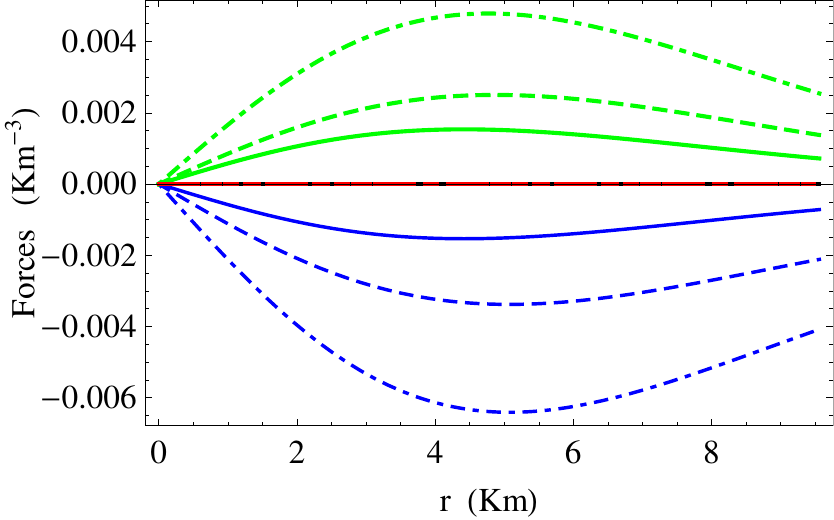}
\caption{Variation of different forces with radial distance $r$~$(km)$ for VELA X-1. Solid, Dashed and Dot Dashed line represent for $D=4$, $D=5$ and $D=6$ respectively. Blue line for $F_{g}$, Green line for $F_{h}$, Red line for $F_{a}$.}
\label{fig23}
\end{figure}

\begin{figure}[h!]
\centering
\includegraphics[width=8.3cm]{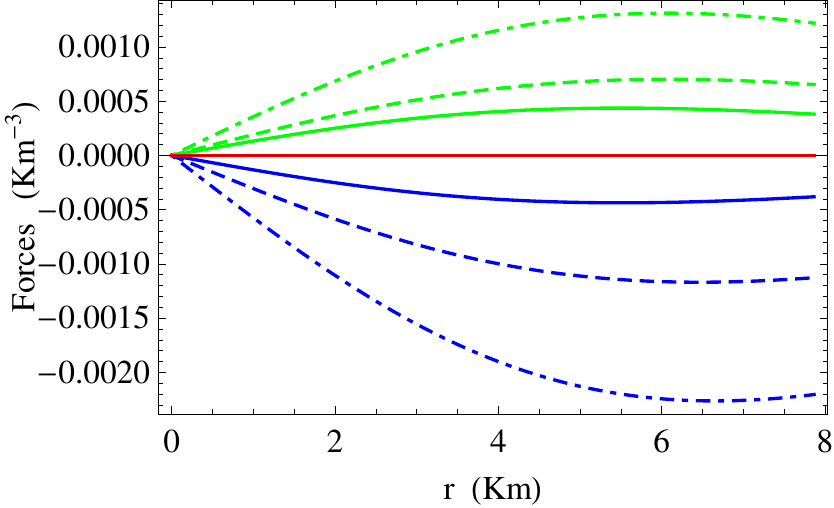}
\caption{Variation of different forces with radial distance $r$~$(km)$ for 4U 1538-52. Solid, Dashed and Dot Dashed line represent for $D=4$, $D=5$ and $D=6$ respectively. Blue line for $F_{g}$, Green line for $F_{h}$, Red line for $F_{a}$.}
\label{fig24}
\end{figure}

\subsection{\textbf{TOV equation}}
The generalized TOV equation \cite{{Tolman},{Oppenheimer}} for the stellar structure is given as
\begin{equation}
-\frac{M_{G}(r)(\rho+p_{r})}{r^2}e^{(\mu-\nu)}-\frac{dp_{r}}{dr}+\frac{2\Delta}{r}=0,\label{eq23}
\end{equation}
where $\Delta$ is the maesure of pressure anisotropy $(p_{t}-p_{r})$, $\mu$, $\nu$ are metric potentials and $M_{G}(r)$ is the gravitational mass contained within the sphere of radius $r$ and the expression of $M_{G}(r)$ is given by
\begin{equation}
M_{G}(r)=r^2\nu^{\prime}e^{(\nu-\mu)}.\label{eq24}
\end{equation}
Equation (\ref{eq24}) can be derived from Tolman-Whittaker mass formula and Einstein field equations. Now substituting eq. (\ref{eq24}) into eq. (\ref{eq23}), we obtain the equation given below:
\begin{equation}
-\nu^{\prime}(\rho+p_{r})-\frac{dp}{dr}+\frac{2\Delta}{r}=0.\label{eq25}
\end{equation}
The TOV equation as given by eq. (\ref{eq25}) has three different parts which are gravitational force $(F_{g})=-\nu^{\prime}(\rho+p_{r})$, hydrostatic force $(F_{h})=-(\frac{dp}{dr})$ and anisotropic force $(F_{a})=\frac{2\Delta}{r}$. In term of these three forces the TOV equation in hydrostatic equillibrium can be written as 
\begin{equation}
F_{g}+F_{h}+F_{a}=0.\label{eq26}
\end{equation}
Eq. (\ref{eq26}) describes the equilibrium of stellar configuration under the combined effect of three different forces namely gravitational force $(F_{g})$, hydrostatic force $(F_{h})$ and anisotropic force $(F_{a})$. The radial variation of $F_{g}$, $F_{h}$ and $F_{a}$ are shown in figure (\ref{fig23}) and (\ref{fig24}) for two different compact objects VELA X-1 and 4U 1538-52 respectively. It is observed that TOV equation holds good under the combined effect of three different forces given above in static equilibrium. It is also observed that gravitational force $(F_{g})$ dominates over  hydrostatic $(F_{h})$ and anisotropic $(F_{a})$ forces but sum of these three forces is always zero for stable stellar structure. 

\subsection{\textbf{Herrrea cracking condition}}
\begin{figure}[h!]
\centering
\includegraphics[width=8.3cm]{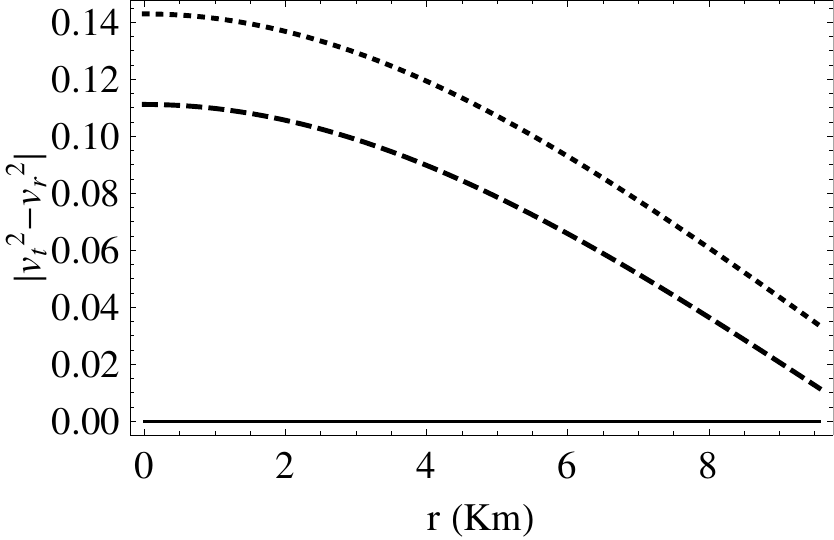}
\caption{Variation of $(v_{t}^2-v_{r}^2)$ with radial distance $r$~$(km)$ for VELA X-1. Solid line for $D=4$, Dashed line for $D=5$ and Dotted line for $D=6$.}
\label{fig25}
\end{figure}

\begin{figure}[h!]
\centering
\includegraphics[width=8.3cm]{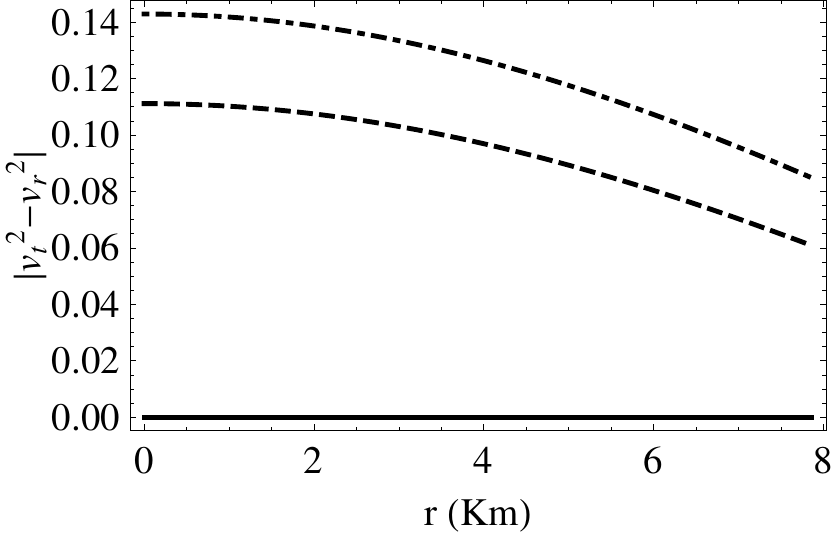}
\caption{Variation of $(v_{t}^2-v_{r}^2)$ with radial distance $r$~$(km)$ for 4U 1538-52. Solid line for $D=4$, Dashed line for $D=5$ and Dot Dashed line for $D=6$.}
\label{fig26}
\end{figure}

The stellar structure of any anisotropic matter configuration of compact objects should be stable and stability can be checked by using the concept of cracking condition introduced by Herrera \cite{Herrera}. Abreu \cite{Abreu} gave an criteria for the stability of stellar configuration on the basis of Herrar's concept. According to Abreu any stellar model is found to be stable if the radial $(v_{r})$ and transverse $(v_{t})$ sound speed satisfy the following condition:
\begin{center}
\begin{equation}
0 \leq \mid v^2_{t}-v^2_{r}\mid \leq 1.\label{eq27}
\end{equation}
\end{center}
Variations of $(v^2_{t}-v^2_{r})$ with radial distance has been graphically shown in figures (\ref{fig25}) and (\ref{fig26}) for two compact objects VELA X-1 and 4U 1538-52 respectively. From figures (\ref{fig25}) and (\ref{fig26}), it is observed that Abreu's inequality condition givn in equation (\ref{eq27}) is satisfied.

\subsection{\textbf{Adiabatic index} }
\begin{figure}[h!]
\centering
\includegraphics[width=8.3cm]{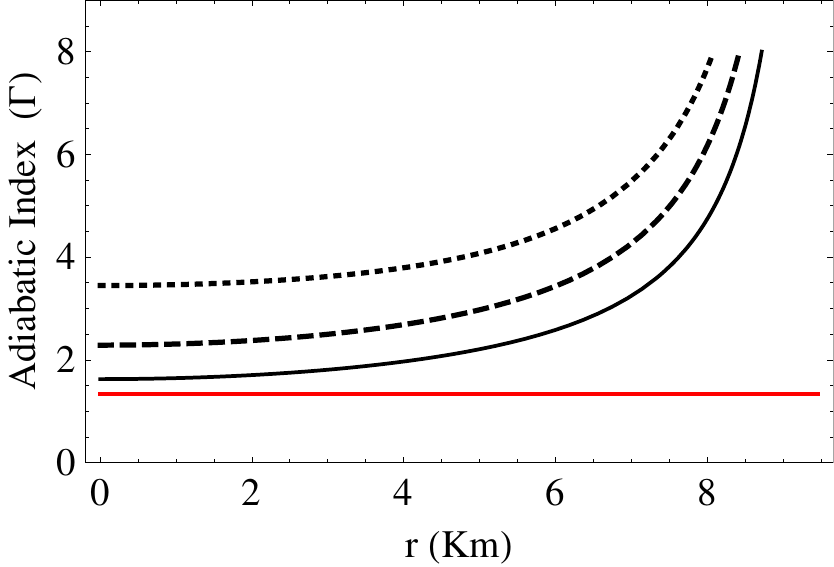}
\caption{Variation of adiabatic index $(\Gamma)$ with radial distance $r$~$(km)$ for VELA X-1. Black solid line for $D=4$, Dashed line for $D=5$ and Dot Dashed line for $D=6$. Red solid line is representing $\Gamma=\frac{4}{3}$.}
\label{fig27}
\end{figure}

\begin{figure}[h!]
\centering
\includegraphics[width=8.3cm]{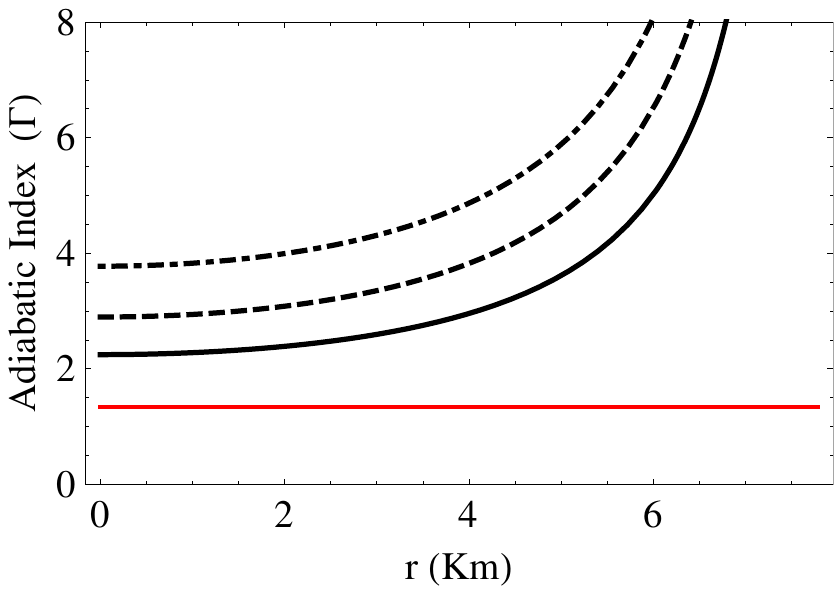}
\caption{Variation of adiabatic index $(\Gamma)$ with radial distance $r$~$(km)$ for 4U 1538-52. Black solid line for $D=4$, Dashed line for $D=5$ and Dot Dashed line for $D=6$. Red solid line is representing $\Gamma=\frac{4}{3}$.}
\label{fig28}
\end{figure}

The stability creterion of a compat star can be studied from the idea of adiabatic index $(\Gamma)$ of the interior matter as described by Heintzmann and Hillebrandt \cite{Heintzmann} and is given below;
\begin{equation}
\Gamma= \frac{\rho+p_{r}}{p_{r}}\frac{dp}{d\rho}=\frac{\rho+p_{r}}{p_{r}}v^2_{r}\label{eq28}
\end{equation}
Heintzmann and Hillebrandt \cite{Heintzmann} established that adiabatic index $\Gamma$ should be greater $\frac{4}{3}$ for a stable stellar configuration. We have evaluated the values of $\Gamma$ from eq. (\ref{eq27}) for two different compact stars VELA X and 4U 1538-52 and are shown in the figures (\ref{fig27}) and (\ref{fig28}) respectively. From figures (\ref{fig27}) and (\ref{fig28}), it is found that the value of $\Gamma$ always greater than $\frac{4}{3}$ i.e. Heintzmann and Hillebrandt condition is satisfied at all points from centre to surface of the compact objects. 

\section{Variation of Lagrangian perturbation of radial pressure with frequency}
\begin{figure}
\centering
\includegraphics[width=8.3cm]{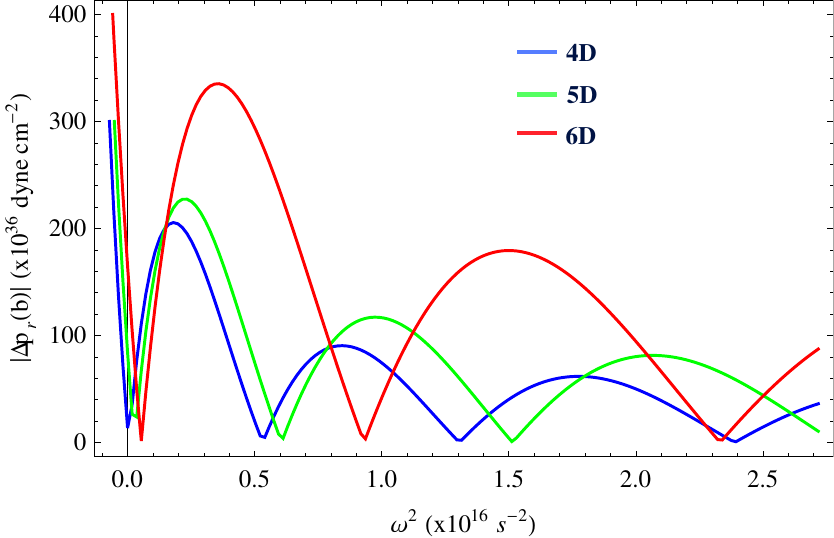}
\caption{Variation of absolute value of Lagrangian change in radial pressure $(p_r)$ with frequency $(\omega^2)$ at the surface of the star for $B= 57.55 MeV/ fm^3$ for VELA X-1 having mass $M=1.77~M_\odot$ and radius $b=9.56~km$.}
\label{fig30}
\end{figure}

\begin{figure}
\centering
\includegraphics[width=8.3cm]{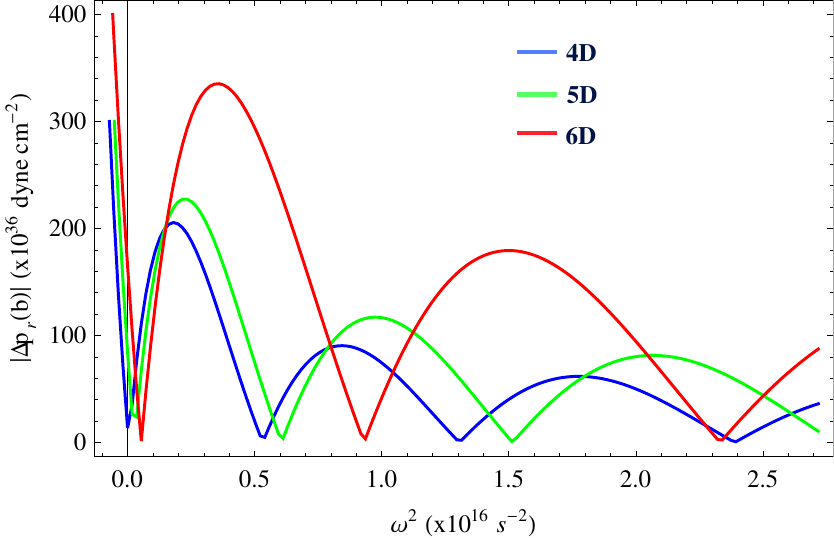}
\caption{Variation of absolute value of Lagrangian change in radial pressure $(p_r)$ with frequency $(\omega^2)$ at the surface of the star for $B= 57.55 MeV/ fm^3$ for 4U 1538-52 having mass $M=0.87~M_\odot$ and radius $b=7.86~km$.}
\label{fig31}
\end{figure}

We also study the variation of Lagrangian perturbation of radial pressure at the surface with respect to frequency ($\omega^2$) for better understanding of the stability of the system against small radial oscillation. In this context, we follow the procedure adopted by Pretel \cite{Pretel}. The equation governing infinitesimal radial mode oscillation can be described in terms of two coupled equations given by
\begin{equation}
\frac{d\chi}{dr}=-\frac{1}{r}\left(3\chi+\frac{\Delta p_r}{\Gamma p_r}\right)+\frac{d\nu}{dr}\chi, \label{eq29a}
\end{equation}

and 

\begin{eqnarray}
\frac{d(\Delta p_r)}{dr}=\chi \left[\frac{\omega^2}{c^2}e^{2(\mu-\nu)}(\rho+p_r)r-4\frac{dp_r}{dr}- \right. \nonumber \\
\left. \frac{8\pi G}{c^4}(\rho+p_r)e^{2\mu}r p_r +r(\rho+p_r)\left(\frac{d\nu}{dr}\right)^2\right] \nonumber \\
-\Delta p_r\left[\frac{d\nu}{dr}+\frac{4\pi G}{c^4}(\rho+p_r)re^{2\mu}\right], \label{eq29b}
\end{eqnarray}
where the eigenfunction $\chi$ is related to the radial part of the Lagrangian displacement by the relation $\chi=\frac{\delta(r)}{r}$. We adopt a simple approach in which the eigenfunctions $\chi$ are normalized so that $\chi(0)=1$ at the centre. In addition it is evident that eq. (\ref{eq29a}) has a singularity at $r=0$. Therefore it is required that the term which is a factor of ($\frac{1}{r}$) must vanish as $r\rightarrow 0$ which gives the condition
\begin{equation}
\Delta p_r = -3\mu \chi p_r \;\; as\;\; r\rightarrow 0. \label{eq29c}
\end{equation} 
Apart from that at the surface ($r=b$) of a star where radial pressure ($p_r(b)=0$) equal to zero, the Lagrangian perturbation of the radial pressure vanishes i.e.
\begin{equation}
\Delta p_r=0 \;\; as \;\; r\rightarrow b. \label{eq29d}
\end{equation}
Here we take a surface density $1.02~\times 10^{14}$~$gm/cm^3$ in $D=4,~5,~6$ dimensions and solve coupled eqs. (\ref{eq29a}) and  (\ref{eq29b}) for different combinations of $\omega$ using the boundary condition given by \ref{eq29c} and $\chi(0)=1$. The absolute value of the Lagrangian change in radial pressure at the surface $|\Delta p_r(b)|$ is plotted against $\omega^2$ as shown in figures \ref{fig30} and \ref{fig31} for VELA X-1 and 4U 1538-52 respectively. The minima of these plots correspond to correct value of the normal mode frequency. It is evident from the figures \ref{fig30} and \ref{fig31} that for all normal modes $\omega_{n}^2>0$. Thus our model is stable against small radial oscillations.

\begin{figure}
\centering
\includegraphics[width=8.3cm]{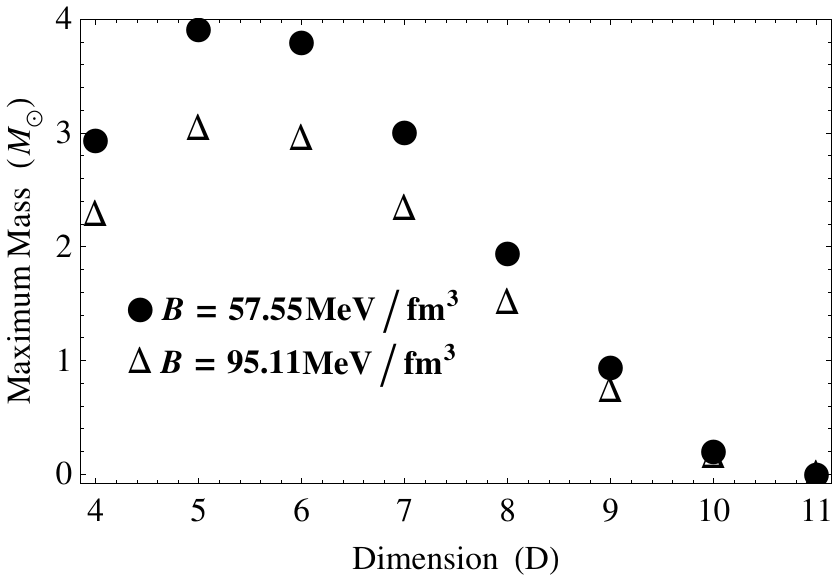}
\caption{Variation of maximum mass $M_{max}~(M_{\odot})$ with dimensions $(D)$. Circular and triangular markers represent respectively for $B=57.55$ and $95.11~MeV/fm^3$.}
\label{fig29}
\end{figure}

\section{Discussion}
We present a class of strange star in higher dimensional Finch-Skea geometry \cite{FS}. We consider Finch-Skea (FS) metric to describe the interior geometry of physical 3-space of a compact star characterized by three constants $C$, $F$ and $S$. Physically viable relativistic stellar models may be obtained using the prescribed metric. To study some physical properties of strange stars, we choose the EOS as $p=\frac{1}{3}(\rho-4B)$ of the interior fluid composed of strange matter as a possible tool. The value of bag constant ($B$) is restricted in the range of $57.55 MeV/fm^3 \leq B \leq 95.11 MeV/fm^3$ required for stable strange matter relative to neutron at zero external pressure \cite{Madsen}. To begin with we choose surface energy density ($\rho_{s}$) of the strange star corresponding to a given value of bag constant $B$ at the surface of the star and found that there exists an upper limit on maximum radius ($b_{max}$) for which the constant $C$ in equation (\ref{eq08}) picks up two real values. Out of these two values of $C$, only one value admit physically viable stellar model satisfying all the necessary physical conditions in four and higher dimensions. After the evaluation of $b_{max}$ numerically, we have determined the maximum mass ($M_{max}$) and compactness ($u_{max}$) of strange star in dimensions $D\ge4$. We have shown the dependence of mass $M$ on $B$ in figures \ref{fig3} - \ref{fig4} for different radii and when $B$ lies within the range of $57.55~\leq~B~(MeV/fm^3)~\leq~95.11$ in four and five dimensions respectively. From figures \ref{fig3} and \ref{fig4}, we note that when radius ($b$) and dimension ($D$) are fixed at certain value, the mass of strange star increases with $B$. Mass-radius relations are shown in figures \ref{fig5} and \ref{fig6} for $B=57.55$ and $B=95.11~MeV/fm^3$ respectively in dimensions $D\ge4$. From figures \ref{fig5} and \ref{fig6}, it appears that the mass ($M$) in four dimensions is higher than in higher dimensions for fixed value of radius ($b$) and bag constant ($B$). We have tabulated the values of maximum radius ($b_{max}$) and corresponding maximum mass ($M_{max}$) and surface red shift $(Z_s)_{max}$ in tables \ref{tab3} and \ref{tab4} for space-time dimensions $D=4$, $5$, $6$ when  $B=57.55~MeV/fm^3$ and $97.11~MeV/fm^3$. Here we have determined the maximum mass of compact objects using the procedure adopted by Sharma et al. \cite{RSharma} and Paul et al. \cite{BCPAUL}. It is observed that when $b_{max}$ and dimensions ($D$) held fixed at certain value maximum mass ($M_{max}$) decreases with the increase of $B$. This type of variation of $M_{max}$ Vs $B$ is explained from different points of view by Harko and Cheng \cite{Harko}. Dependence of maximum mass ($M_{max}$) of strange stars on the space-time dimensions ($D$) are given in figure (\ref{fig29}). It is found that maximum mass ($M_{max}$) picks up maximum value at $D=5$ when $B$ is fixed and decreases when $D=4$ and $D\ge6$. This type of variation has also been predicted by Paul et al \cite{BCP} considering different metric. From figure (\ref{fig29}), we note one interesting result that when space-time dimensions ($D$) increases the effect of $B$ on the value of maximum mass ($M_{max}$) decreases and the effect is almost negligible when $D\ge11$. From tables \ref{tab3} and \ref{tab4}, it is observed that space-time dimensions has some effects on maximum compactness ($u_{max}$). $u_{max}$ picks up lower value in higher space-time dimensions ($D$). Strange star having compactness greater than 0.3 are supposed to be Type-I Strange Star as prescribed by Tikekar and Jotania \cite{Jotania}. From tables \ref{tab3} and \ref{tab4}, we note that maximum compactness in four and five dimensions are above $\ge0.3$. The maximum surface redshift ($(Z_{s})_{max}$) is evaluated in this model and is tabulated in table \ref{tab3} and \ref{tab4} for $B=57.55 MeV/fm^3$ and $B=95.11 MeV/fm^3$ respectively in $D=4$, $D=5$ and $D=6$. We note that $(Z_{s})_{max}$ is maximum at $D=4$ and decreases with increase of $D$. It is also noted that $(Z_{s})_{max}$ changes appreciably with the change in $B$ keeping $D$ constant. In this model, we found the value of maximum surface redshift as $0.8822$ when $B=57.55~MeV/fm^3$ in four dimension. Buchdahl \cite{Buchdahl} predicted that maximum surface red shift for any perfect fluid distributed in a spherical region have the value $(Z_{s})_{max}\le2$. Bohmer \cite{Bohmer} later generalized the work of Buchdahl and pointed out that maximum surface red shift may be higher than the previously obtained limit $2$. According to Bohmer \cite{Bohmer} $(Z_{s})_{max}\le5$. Maximum surface redshift as predicted by Bohmer \cite{Bohmer} is consistent with the work of Ivanov \cite{Ivanov} having the value $(Z_{s})_{max}\le5.211$. We note the value of maximum surface redshift obtained in this model consistent with these allowed limits. All the energy conditions for a physically viable stellar model are described in figures ~\ref{fig7} - \ref{fig12}. We note that in the present model all the necessary energy conditions hold good for the chosen parameters. From figures \ref{fig13} - \ref{fig14}, we conclude that causality conditions hold good in the present model. In this model, we observe that maximum radius and maximum mass of strange star may vary in the range $9.41-12.10~km$ and $2.29~M_{\odot}$ to $2.94~M_{\odot}$ respectively in four dimensions when $B$ varies from $95.11 MeV/fm^3$ to $57.55 MeV/fm^3$ respectively. But in $D=5$ dimensions, the values of maximum radius and maximum mass range between $13.77-17.70~km$ and $3.05~M_{\odot}$ to $3.91~M_{\odot}$ when $B$ changes from $95.11 MeV/fm^3$ to $57.55 MeV/fm^3$ respectively. Therefore physically viable strange star configuration may be predicted by adjusting bag constant $B$ and dimensions $D$. We notice that maximum compactness always lies below the limit as predicted by Buchdahl \cite{Buchdahl} and Ponce De leon \cite{PDL}. We have applied our model to study physical properties of two star VELA X-1 and 4U 1538-52 which are supposed to be strange stars. Radial variation of energy density ($\rho$), radial ($p_{r}$), transverse ($p_{t}$) pressures and anisotropy ($\Delta$) for VELA X-1 are shown in figures \ref{fig15}-\ref{fig18} respectively. Radial variation of energy density ($\rho$), radial ($p_{r}$), transverse ($p_{t}$) pressures and anisotropy ($\Delta$) for 4U 1538-52 are shown in figures \ref{fig19}-\ref{fig22} respectively. From the figures \ref{fig15}-\ref{fig22}, we found that when dimension increases the value of $\rho$, $p_{r}$, $p_{t}$ and $\Delta$ also increase. In this model, it may be possible to predict the observed mass of a strange star when observed radius is taken as input parameter. Here in table~\ref{tab5}, we have predicted the mass $1.77~M_{\odot}$ of VELA X-1 in different dimensions for specific value of $B$ when its observed radius is fixed at $9.56~km$. Same prediction for the observed mass $0.87~M_{\odot}$ of 4U 1538-52 is possible for the observed radius $7.866~km$ and are tabulated in table~\ref{tab7}. We observe that the maximum mass $3.2~M_{\odot}$ as predicted by Rhoades and Ruffini \cite{Rhoades} may be obtained for $D=5$ and $B=93.81~MeV/fm^3$. The stability criterion is studied from the following analysis (i) generalized TOV equation, (ii) Herrera cracking condition and (iii) Adiabatic index.  The results are shown graphically in figures \ref{fig23}-\ref{fig28} for two stars namely VELA X-1 and 4U 1538-52. Figures ensure that present model satisfies all stability criterion. We have also studied the variation of Lagrangian perturbation of radial pressure with frequency. The absolute value of the Lagrangian change in radial pressure at the surface $|\Delta p_r(b)|$ of the star is plotted against $\omega^2$ as shown in figures \ref{fig30} and \ref{fig31} for VELA X-1 and 4U 1538-52 respectively. The minima of these plots correspond to correct value of the normal mode frequency. It is evident from the figures \ref{fig30} and \ref{fig31} that for all normal modes $\omega_{n}^2>0$. Thus our model is stable against small radial oscillations also.

\section*{Acknowledgement}
BD and KBG are thankful to CSIR for providing the fellowship vide no 09/1219(0005)/2019-EMR-I and 09 /1219(0004)/2019-EMR-I respectively. AS is thankful to Physics Dept. CBPBU for extending the facilities to carry out research work.

\begin{description}
\item[Availability of data and material:] 
This manuscript has no associated data or the data will not be deposited, we have used only observed mass and radius of some known compact objects to construct relativistic stellar models. 
\end{description}

\end{document}